\pdfoutput=1

\documentclass[sigconf,screen]{acmart}
\AtBeginDocument{%
  \providecommand\BibTeX{{%
    \normalfont B\kern-0.5em{\scshape i\kern-0.25em b}\kern-0.8em\TeX}}}

\copyrightyear{2025}
\acmYear{2025}
\setcopyright{rightsretained}
\acmConference[FSE Companion '25]{33rd ACM International Conference on the
Foundations of Software Engineering}{June 23--28, 2025}{Trondheim, Norway}
\acmBooktitle{33rd ACM International Conference on the Foundations of
Software Engineering (FSE Companion '25), June 23--28, 2025, Trondheim,
Norway}\acmDOI{10.1145/3696630.3734199}
\acmISBN{979-8-4007-1276-0/2025/06}

\usepackage{tabularx}
\newtheorem{dfn}{Definition}
\newtheorem{ex}{Example}

\newcommand{\parent}[1] {\textsf{Parent}(#1) }

\newcommand{\builds}[2] {\textsf{Builds}_{#1}(#2) }
\newcommand{\passes}[2] {\textsf{Passes}_{#1}(#2) }
\newcommand{\fails}[2] {\textsf{Fails}_{#1}(#2) }
\newcommand{\oracle}[2] {\textsf{Oracle}_{#1}(#2) }
\newcommand{\hardening}[2] {\textsf{Hardening}_{#1}(#2) }

\newcommand{\Shardening}[2] {\textsf{SHardening}_{#1}(#2) }
\newcommand{\PPhardening}[2] {\textsf{PPHardening}_{#1}(#2) }
\newcommand{\PRhardening}[2] {\textsf{PRHardening}_{#1}(#2) }
\newcommand{\catching}[2] {\textsf{Catching}_{#1}(#2) }
\newcommand{\Scatching}[2] {\textsf{SCatching}_{#1}(#2) }
\newcommand{\Rcatching}[2] {\textsf{RCatching}_{#1}(#2) }
\newcommand{\Fcatching}[2] {\textsf{FCatching}_{#1}(#2) }
\newcommand{\SRcatching}[2] {\textsf{SRCatching}_{#1}(#2) }
\newcommand{\SFcatching}[2] {\textsf{SFCatching}_{#1}(#2) }

\begin{document}

\title{Harden and Catch for Just-in-Time Assured LLM-Based Software Testing: Open Research Challenges}


\author{Mark Harman}
\orcid{https://orcid.org/0000-0002-5864-4488}
\affiliation{%
  \institution{Product Compliance and Privacy team, Meta Platforms}
  \city{London}
  \country{UK}
}
\affiliation{%
  \institution{University College London}
  \city{London}
  \country{United Kingdom}
}

\author{Peter O'Hearn}
\orcid{0009-0007-0910-5350}
\affiliation{%
  \institution{Fundamental AI Research team, Meta Platforms}
  \city{London}
  \country{UK}
}
\affiliation{%
  \institution{University College London}
  \city{London}
  \country{United Kingdom}
}

\author{Shubho Sengupta}
\orcid{https://orcid.org/0009-0007-4204-5185}
\affiliation{%
  \institution{Independent Software and AI Consultant}
  \country{USA}
}

\renewcommand{\shortauthors}{Harman, O'Hearn, Sengupta}

\begin{abstract}
Despite decades of research and practice in automated software testing, several fundamental concepts remain ill-defined and under-explored, yet offer enormous potential real-world impact.
We show that these concepts raise exciting new challenges in the context of Large Language Models for software test generation.
More specifically, we formally define and investigate the properties of hardening and catching tests.
A hardening test is one that seeks to protect against future regressions, 
while a catching test is one that catches such a regression
or a fault in new functionality introduced by a code change.
Hardening tests can be generated at any time and may become catching tests when a future regression is caught.
We also define and motivate the Catching `Just-in-Time' (JiTTest) Challenge, in which tests are generated `just-in-time' to catch new faults before they land into production. 
We show that any solution to Catching JiTTest generation can also be repurposed to catch latent faults in legacy code.
We enumerate possible outcomes for hardening and catching tests and JiTTests, 
and discuss open research problems, 
deployment options, 
and initial results from our work on automated LLM-based hardening at Meta.
This paper\footnote{Author order is alphabetical. The corresponding author is Mark Harman.} was written to accompany the keynote by the authors at the ACM International Conference on the Foundations of Software Engineering (FSE) 2025.
\end{abstract}

\ccsdesc[500]{Software and its engineering~Software testing and debugging}
\keywords{Software Engineering, Programming Languages, Large Language Models (LLMs). }

\maketitle

\section{Introduction}
For the past decade, at Meta, we (the authors) have been developing automated static and dynamic analysis and testing technologies. 
We have also been involved in the development of AI technologies, as a support to such analyses, and also as an independent technology in its own right.
As well as our recent industrial experience, we have been active in the Software Engineering, Programming Languages, and Artificial Intelligence research communities over several decades.
Our observation is that, despite much research and development, some of the most fundamental concepts that lie at the intersection between scientific research and practical software testing have yet to be adequately analysed, formalised, or defined, let alone fully tackled.

Research and practice are unequivocally and rapidly moving towards automated software test design, 
increasingly relying on language models to generate the test cases~\cite{mhetal:LLM-survey}.
This makes it more important than ever to have crystal clear definitions of the most pressing problems facing practising software engineers, when it comes to automated test generation~\cite{anandetal:orchestrated}.
Automated test generation is not only important for testing code generated, both by humans and models, it may also prove useful for training and assessing the models themselves \cite{jain2024testgeneval}.

In this paper, we attempt to provide simple, intuitive, and precise formulations of the underlying concepts on which automated software test design rests.
We first formalise the way in which generated tests harden against future regressions. 
This {\em regression} test generation has been the primary focus of both academic research and practice in automated test generation for the past few decades, with notable successes~\cite{mh:icst15-keynote,cadar:three-decades,mhetal:ssbse18-keynote, zalewski:afl}.
However, as we show, it leads us into a Regression Only Trap, in which tests must pass on the current revision, and therefore cannot catch new bugs introduced by that revision, nor long-standing legacy bugs.

These regression tests are `hardening' tests according to our terminology, because they seek to detect bugs in future changes rather than in the current system; they `harden' against future regressions.
Research and practice has tended to focus on regression testing due to  the Oracle Problem~\cite{ebetal:oracle}.
Regression testing offers the automated test tool designer a readily deployable solution to the Oracle Problem:
the Regression Oracle~\cite{alshahwan:software}.
The Regression Oracle dictates that the behaviour of the previous version of the system should be preservered by any change, thereby ensuring that there are no regressions.

In Section~\ref{sec:atMeta} we briefly review our previous work on automated test generation at Meta, 
leading up to our current work on
LLM-based automated unit test generation for hardening.
The primary contributions of this work lie in the way we use LLMs and mutation.

Nevertheless, by remaining stuck in the  Regression Only Trap, we (and the wider the community), have overlooked
an important untackled problem: the Catching Just-in-Time test (Catching JiTTest)  problem.
A  JiTTest is generated when a pull request is submitted.
We also formalise and analyse JiTTests.
By seeking to generate a catching JiTTest, we aim to catch bugs specifically {\em in} that pull request.
A JiTTest is not concerned with some possibly distant future in which some regression may (or may not) occur, 
but the {\em immediate} future in which a pull request is about to land into production.

Generating Catching JiTTests allows us to escape the Regression Only Trap (ROT), 
but it represents a more challenging problem than generating hardening regression tests, although it subsumes that problem.
It is also much more impactful. 
A JiTTest can catch bugs in new functionality {\em as well as} regressions, and it stops them literally {\em just-in-time}. 
That is, just before the pull request lands into production.
Because they do not merely target regressions,  JiTTests can {\em also} 
easily be repurposed to find {\em latent} bugs in the {\em existing} code base, 
as we show in Section~\ref{sec:latent}.
We believe the Catching JiTTest Challenge to be the most challenging and impactful problem in software testing.
Although there has been initial discussion of just-in-time testing approaches ~\cite{fernandez1996using,veanes2005fly,mossige2015testing,brandt2022developer}, much more remains to be done.

Section~\ref{sec:formal}, introduces our formalisation of hardening and catching. 
Our formalism allows us to precisely determine when a hardening test subsequently catches a bug.
It also allows us to distinguish between tests that catch regressions and those that catch bugs in new functionality.
We use our formalism to precisely define the Catching JiTTest Challenge.
Our definitions apply to testing in general, whether tests are generated by machine or by human, 
and irrespective of the generation technique used. 
Nevertheless, the motivation for our work is our recent development and deployment \cite{mhetal:TestGen-LLM,foster:mutation} 
of LLM-based test generation (using an Assured LLMSE approach~\cite{mhetal:intense24-keynote}). 

The paper also includes a case-based analysis of possible deployments for hardening and catching tests, considering the balance of risk and reward inherent in their deployment.
This analysis reveals new insights into deployment options.

The primary contributions of this paper are as follows:

\begin{enumerate}
    \item {\bf Introduction of the Catching JiTTest Challenge}:
    Although challenging, we argue that the Catching JiTTest Challenge is the most highly impactful currently open challenge in automated test generation. 
    We also show how advances in LLMs open the door to previously unavailable solutions, 
    making the time ripe to tackle Catching JiTTest Challenge. 

    \item {\bf Precise Formal Foundations}: A formalisation of hardening and catching tests, and the distinction between regression catching and functionality catching tests. 
    
    \item {\bf Detailed Practical Deployment Analysis}: A case-by-case analysis of deployment options, illustrating how the formalism informs practical deployment options.
\end{enumerate}

\section{Automated Testing at Meta}
\label{sec:atMeta}
In this section, we briefly review our previous and current work, at Meta, 
on automated test generation.
The section is a brief review of work published in more detail elsewhere.
The aim of the section is to make the paper self-contained and to motivate the need for the more detailed analysis of hardening and catching tests, 
introduced in Section~\ref{sec:formal}.

\subsection{Background}
At Meta, we have previously deployed end-to-end  test generation as well as unit-level test generation.
Our end-to-end testing tools include
Sapienz~\cite{mao:sapienz:16}, deployed since 2017~\cite{mhetal:ssbse18-keynote} covering client-side testing on all platforms, 
and Fausta, deployed since 2021~\cite{kmetal:fausta}, covering server-side testing for WhatsApp.
The initial Sapienz deployment in 2017 used the implicit oracle~\cite{ebetal:oracle}, 
which has remained widespread throughout the company since 2017.
More recently, we enhanced the Sapienz oracle to extend beyond the implicit oracle~\cite{tuli:simulation}, 
and to cater  for the rich diversity of user states needed to cover the state space ~\cite{kinga:enhancing}.

These end-to-end testing systems can be thought of as simulation-based testing approaches~\cite{jaetal:gi20-keynote},
for which we have used Metamorphic Testing to test the simulation itself~\cite{jaetal:mia}.
In traditional end-to-end testing, a test case simulates the end-to-end behaviour of the platform with respect to a single user.
We have also deployed simulation-based  
test generation techniques that simulate the end-to-end behaviour of an entire interacting community of users.
We call this approach `social testing'. 

Social testing is not only applicable to any platform in which users interact with each other, it is also necessary so that these interactions are fully tested~\cite{tuli:simulation,jaetal:ease21-keynote}.
As we have shown in our previous work, social testing goes beyond traditional end-to-end testing, 
because it takes into account user `personas', 
and the fact that the test oracle for one user differs (and may even {\em contradict}) the test oracle for another user.

These dynamic analyses are complemented by static analyses ~\cite{distefano:scaling}, such as
Infer~\cite{movefast} and Zoncolan.

In addition to end-to-end test generation, we have also more recently developed unit-level test generation techniques using language models~\cite{mhetal:TestGen-LLM} and observations~\cite{mhetal:TestGen-obs}.
These unit test generation techniques focus on covering uncovered code.

To go beyond merely achieving additional structural code coverage, we also developed a mutation-guided approach~\cite{foster:mutation}.
The mutation-guided approach ensures that the generated tests find faults that no other test can find.
This more recent mutation-guided approach generates hardening tests (according to Definition~\ref{dfn:weak-hardening}), 
because the tests are {\em guaranteed} to find at least one currently uncaught potential future fault.

\subsection{Assured Large Language Model Based Software Testing (Assured LLMST)}
\label{sec:assured}
Our approach to the generation of hardening tests is grounded in the principles of
Assured LLM-based Software Testing (Assured LLMST).
Assured LLMST is simply
Assured LLM-Based Software Engineering (Assured LLMSE) applied to the field of Software Testing.

Assured LLMSE ~\cite{mhetal:intense24-keynote} is a generate-and-test approach to LLM-based software generation, inspired by the Genetic Improvement~\cite{Petke:gisurvey} approach to Search Based Software Engineering~\cite{mhbj:manifesto,mhamyz:acm-surveys,ramirez2019survey}, also widely used in Software Testing~\cite{mcminn:survey,mh:icst15-keynote}.
Assured LLMST is for Assured LLMSE, what Search Based Software Testing (SBST) is for Search Based Software Engineering (SBSE); it focusses on that subset of Software Engineering that is concerned  with Software Testing.

Assured LLMSE addresses the  two fundamental questions:
How can we provide verifiable assurances that the LLM-generated code 
\begin{enumerate}
\item does not regress the properties of the original code ?
\item improves the original in a verifiable and measurable way ?
\end{enumerate}

\subsection{Assured LLMST is Easier than Assured LLMSE}
Fortunately Assured LLMS{\em T} is considerably less challenging, according to both these constraints, 
then LLMS{\em E}.

\noindent
{\bf Does not Regress}:
In the context of software testing, not regressing the properties of the original code means not regressing the properties of the original test suite.
For a test suite, the typical properties of interest lie in its
ability to achieve a certain level of coverage and
to catch a certain class of faults.
When we add new test cases to an existing test suite (but do not remove any), 
we cannot, by definition, regress the test suite's ability to achieve coverage, nor its ability to reveal faults.
Provided we treat each test case as a unique discrete unit (on which existing tests have no dependency), then
adding additional test cases can only improve the overall test suite.
This is in stark contrast to the situation in which we use language models to generate code {\em other} than tests.
When generating code in general, it is easy to accidentally regress existing functionality.

\noindent
{\bf Improves the Original}:
For the application to software testing, improving the original in a verifiable and measurable way is also simple:
We can measure the additional coverage and/or can give an assurance that the new test finds a fault that is uncaught by any existing test.
By contrast, the goal of improving code, {\em other} than test code, in a verifiable  and measurable way can be challenging.
For example, attempting to improve code execution time  raises the challenges of measuring execution time reliably, over a distribution of potential inputs.

\subsection{Online Assured LLMST}
\label{sec:online}
As with all machine learning~\cite{hoi:online}, LLMSE can be either `online' or `offline'~\cite{mhetal:intense24-keynote}.
The extra time available to offline LLMSE typically allows the Assured LLMSE application
greater opportunities to provide assurances.

One might naturally think of an `online' approach as one that responds in {\em real} time.
The archetype of online LLMSE is code completion.
Code completion technologies, such as CoPilot~\cite{copilot2023} and CodeCompose~\cite{murali:codecompose},
 provide the generated code directly in the editor as the software engineer types text.
This is clearly online because the response latency is the same as the duration of a keystroke.

When we formalise the notion of online, 
it becomes clear that there is, perhaps surprisingly, 
far more time available for some generated test deployment options than there is for code completion.
The definition of `online' (for the purposes of distinguishing between online and offline LLMSE)
rests on the definition of `real time', which has its own notion of timeliness~\cite{dodhiawala:real,kopetz:real}.
More formally, the LLMSE definition~\cite{mhetal:intense24-keynote} of real time is repeated below as Definition~\ref{dfn:rt}.

\begin{dfn}[Real Time, taken from ref~\cite{mhetal:intense24-keynote}]
\label{dfn:rt}
A result is provided in {\em real time} for a consumer $c$ and application $a$, if and only if $c$ cannot perform $a$ any better when the time spent awaiting the LLM response reduces.
\end{dfn}

This is a `utility'-based definition of real time.
It focusses on the code being provided in {\em sufficient} time to be acted upon.
Any JiTTest generation approach that contributes tests generated on the fly for a pull request
merely has to generate these tests before the first human reviewer considers the pull request.

It may take a human reviewer hours or even days to provide that first review.
According to a recent study~\cite{hasan2023understanding}, the median response time
for Git projects was estimated at 1.75 hours. 
This estimate includes  (instant response) static analysis bots (such as lint tools), which tends to make it an underestimate of the time available to the first human response.
Even with this estimate, any test generated within 1.75 hours would thus be `in time' for the review process with a probability of at least  0.5.

In fact, a more realistic estimate of the typical industry standard time for the first {\em human} response
to a pull request is closer to 8 hours ~\cite{mcgill:time}.
That is, there is an aspiration to respond within eight hours, and this is also therefore a realistic bound for `real time' test generation (and thus for a JiTTest).
As can be seen, the constraints for a test to be considered a JiTTest are not as tight as they might initially appear.
This is important because eight hours affords a great deal of computational time that our tools and infrastructure can use to evaluate the assurances that can be offered to the engineer as part of the review process.

\subsection{Initial Results on LLMST}
At Meta, we recently developed and deployed a  mutation-guided test generation system called ACH~\cite{foster:mutation}.
ACH  ensures that the generated tests are genuinely hardening; they can catch a potential future bug. 
This should be contrasted with test generation that aims to improve structural code coverage but may not necessarily prove to be fault-revealing.
Furthermore, since our deployment places test generation in a scenario in which the mutant is {\em guaranteed} to be killed, we also sidestep the familiar equivalent mutant problem~ \cite{yjmh:analysis,xymhyj:equivalent,papadakis:trivial,van2021mutantbench,madeyski2013overcoming}.

The overall approach allows the ACH tool to give the following six assurances, as part of its overall Assured LLMSE approach.

\begin{enumerate}
    \item {\bf Buildable}: The proposed new tests build, so are free from syntax errors and missing dependencies;
    \item {\bf Valid Regression Tests}:  The tests pass and do so consistently, so they are non-flaky~\cite{mhpoh:scam18-keynote} regression tests;
    \item {\bf Hardening}: The new tests catch faults that {\em no} existing test can catch. 
    Our mutation-guided tests are `hardening' according to Definition~\ref{dfn:weak-hardening}. 
    Furthermore, when an engineer accepts and lands such a generated test into production, the test becomes a Strong Hardening Test according to Definition~\ref{dfn:strong-hardening}.
    \item {\bf Additional Coverage}: The tool automatically reports any additional coverage achieved.
    \item {\bf Relevant}: Many of the new tests are closely coupled to the issue of concern;
    \item {\bf Fashion Following}: Most of the new tests respect the coding style used by the existing tests available for the class under test; they are `fashion followers'~\cite{mhetal:TestGen-LLM}.

\end{enumerate}

The first four of these six assurances are assurances in the strict meaning of the word: they are verifiable (and, therefore, falsifiable) guarantees that leave no room for doubt.
The final two stretch the definition of `assurance' a little bit; they are non-boolean aspirational attempts to provide improvement, in the sense of genetic improvement guarantees~\cite{Petke:gisurvey}.
That is,  we hope that the test will `fashion follow', and therefore be more intuitive to the human reader, but this is something that can be somewhat subjective, rather like genetic improvement and restructuring~\cite{raiha:survey,laurie:gecco07}  or other aesthetic software improvements~\cite{ramirez2018systematic,simons:interactive}.
We also hope the tests will be relevant, but this is also a somewhat subjective judgment.

We  found that tests designed to be hardening (by killing mutants) that were deemed to be irrelevant 
by the engineer reviewing them, were more likely to be accepted when they improved coverage. 
Such tests proved acceptable even when they did not have another contribution~\cite{mhetal:TestGen-LLM,foster:mutation}.
Of course, coverage improvement is not a necessary criterion for a test to be deemed acceptable
because the test may reveal a fault without covering any new lines of code. 

Therefore, we report additional coverage as a spinoff benefit; it is a measurable assurance we can track, but do not target.
In managing software engineering deployment, it is important to distinguish between metrics that merit tracking from those that should become goals and Key Performance Indicators.

More details about our work on LLM-based test extension can be found in our FSE 2024 paper~\cite{mhetal:TestGen-LLM},
while the more recent approach to incorporate mutation testing to guide the generation of tests toward those that harden with respect to specific faults (using mutation) can be found in our FSE 2025 paper~\cite{foster:mutation}.

\section{Formalising Hardening and Regression and Functionality Catching  Tests}
\label{sec:formal}
In this section we formalise the notions of hardening and catching tests using these to subsequently distinguish regression catching and functionality catching tests.
The formalism allows us to make these distinctions more precise.
In Section~\ref{sec:case-by-case} we use it to analyse, case by case, the possible signals we may obtain from tests, the possible underlying causes, 
and consequent courses of action available when deploying hardening and catching test generation.

Software engineers typically deploy their code into a continuous integration pipeline, 
consisting of a sequence of `revisions' (aka `versions' or `pull requests').
We purposely do not over-constrain the concept of a version. 
This could be a version that has been released, or one that is purely internal.
The version may reside on a  `forked' branch or may reside on the `trunk' of the commit tree.
For the purposes of defining Just-in-Time software test generation, 
these details are irrelevant; we simply need a way to refer to the parent  of a given code version.

\begin{dfn}[Fundamentals]
Defintion: We assume 
\begin{enumerate}
\item A tree repository with a function $\parent{R}$ for a revision  (aka pull request) $R$. 
\item A set Test Cases, $T$
\item Predicates $\builds{R}{t}$, $\passes{R}{t}$, $\fails{R}{t}$, for a revision  $R$ and test case $t$.
\end{enumerate}
\end{dfn}

The intuition behind these relations is as follows.
When a test case, $t$ builds on revision $R$ this is denoted $\builds{R}{t}$.
A test case that builds can be executed. 
When executed, a test yields a signal that includes at least\footnote{
There are often other possibilities such as timeouts, which may signify an engineering concern about incorrect test infrastructure  operation. 
We do not want to constrain our definitions to include these signals, since they are engineering details that are irrelevant to understanding the fundamental signal of a correctly operating hardening or catching test.
} 
two possible outcomes, pass and fail.
When the outcome of executing test $t$ on revision $R$ is pass, this is denoted $\passes{R}{t}$.
When the outcome of executing test $t$ on revision $R$ is fail, this is denoted $\fails{R}{t}$.
A Test suite $T$ is simply a set of test cases.
Executing a test suite simply means executing each of the tests in the suite.

We make the simplifying assumption that tests can be executed independently.
That is, the execution of one test from a suite does not influence the outcome of some other test from the suite.
All of our definitions and concepts can be adapted to relax this independence assumption, but it makes the treatment more complex.

At first sight, it might seem that {\em all} tests should/would be buildable, but this is incorrect. 
When a new revision is created, some of the existing tests may become invalid (unbuildable). 
For example, if the name of a method is changed, then any test that relies on the method name will no longer build.
Therefore, for each revision, not all tests will necessarily be buildable.

We are concerned with test timeliness, because we want to define just-in-time test generation.
We therefore define a timely test, but do not require that it be executable, merely that it be available:

\begin{dfn}[Timely Test]
A timely test for the revision $R$, is available for attempted execution on $R$.
That is, the test exists on or before the time at which the revision $R$ is first submitted as a pull request for review. 
\end{dfn}

We want tests to be available when needed, but we also want to define
a specific kind of test that is literally `just-in-time' for a revision $R$:

\begin{dfn}[Just-In-Time Test (JiTTest)]
A Just-in-Time test, or `JiTTest', for revision $R$, is a timely test for $R$ that is not a timely test for $\parent{R}$.
\end{dfn}

A JiTTest is a kind of on-the-fly~\cite{fernandez1996using,veanes2005fly} or just-in-time test~\cite{mossige2015testing,brandt2022developer}, 
in which we seek to construct a test as soon as the revision appears as a pull request.
An engineer could write a JiTTest. 
Indeed, many engineers do write unit tests just-in-time, because they create them to accompany their pull requests.

In this paper, we are particularly interested in generated JiTTests, because this is an application of test generation that has been largely ignored in the research  literature.
The JiTTest may harden the pull request against future regressions (Hardening JiTTest) or may catch new bugs in it (Catching JiTTest).
As we shall see, Catching JiTTest generation  represents the most challenging, but also most impactful application area for automated test generation.
We are also interested in timely hardening test generation.
Although this is less challenging that Catching JiTTest generation, 
it is also useful, and represents the current deployment mode for most automated test generation, 
both in the research and practitioner communities.

We do not attempt to formally define `generated test' for revision $R$,
but our intuition is that a generated test is simply any test that is generated by machine, 
without human supervision.

\subsection{Coverage}
We want to be able to distinguish useful tests from useless (e.g., trivial) tests.
For example, the trivial test {\tt Assert(True)} is guaranteed to be both timely and buildable
for any revision, according to our definitions, but clearly offers no value.

Coverage is currently the most widely used criterion for test generation in both academic research and industrial deployment.
For example, there are automated test data generation systems based on  Search Based Software Testing (SBST)~\cite{mh:icst15-keynote,mcminn:survey}  
such as Austin~\cite{kletal:austin-ist} for C
and EvoSuite~\cite{fraser:evosuite} for Java, 
and test generation systems based on concolic execution~\cite{godefroid:dart,sen:cute,cadar:three-decades} 
such as Klee~\cite{cadar:icse11} 
that are publicly available and have been used both in research work and in practical deployment.
These technologies focus on coverage achieved, seeking to increase coverage by making it a search objective, in the case of SBSE deployment, or making it the focus of a mixture of symbolic and concrete execution in the case of concolic execution systems.

Automated end-to-end test generation systems such as 
Sapienz~\cite{mao:sapienz:16,mhetal:ssbse18-keynote} 
and other simulation-based testing technologies, such as
Meta's WW~\cite{jaetal:ease21-keynote} 
and Fausta ~\cite{kmetal:fausta}, 
and Google's RecSim~\cite{Ie:RecSym} 
also increase coverage, but as a side benefit of seeking to catch bugs.
These end-to-end systems target the detection of bugs rather than the elevation of coverage.
They tend to rely on the so-called implicit oracle~\cite{ebetal:oracle} or the regression oracle~\cite{alshahwan:software}, so they cannot target the revelation of faults in new functionality. 

Automated fuzzing techniques~\cite{manes:fuzzing}, 
such as AFL~\cite{zalewski:afl} have also been proven to be successful in industrial deployment, 
due to their conceptual simplicity, applicability and ability to catch security flaws.
These technologies also tend to focus on increasing the coverage of the system under test and use the implicit oracle,
traditionally focusing on features that, in any and all applications, could lead to a security flaw.

The implicit oracle captures that which no software system should ever do (such as crashing and running out of memory), 
rather than the properties that we expect for a specific system (such as its functional and other behavioural properties).
This means that implicit oracle tests can only target a very restricted, albeit highly egregious, class of bugs.

Sadly, any coverage-improving test may
simply execute code that does not matter while chasing the goal of elevating coverage.
The freshly covered code may not matter because, for example, it is never executed by any users.
Even when a new test covers previously uncovered code that {\em does} matter, it may not be executed in a way that can reveal bugs.
It is well known, from many empirical studies, that increasing coverage does not provide a guarantee of increasing fault revelation~\cite{frankl_etal97,gopinath2014code,inozemtseva2014coverage,just:are-mutants}.
In fact, evidence tends to suggest~\cite{mike:icse17} that the best techniques to increase fault revelation focus specifically on {\em faults} targeted by mutation testing~\cite{yjmh:analysis}.

These empirical observations are also born out by our definitions.
A test can increase coverage, yet be guaranteed, by construction, to never reveal {\em any} bugs.
For example, consider a test that executes additional lines of code yet relies on an oracle (assertion) that is semantically equivalent to {\tt Assert(True)}.
This is a rather extreme case of the well-known coverage dilemma, but it illustrates the fundamental underlying problem. 
To address this, we define a `hardening' test case; one that hardens the code against future changes by catching {\em at least one bug} 
or type of bug in a well-defined way.

\subsection{Oracles}
\label{sec:oracle}
In order to define a hardening test case, we first need to introduce the concept of an oracle~\cite{ebetal:oracle}.
Strictly speaking, we do not require a full oracle,
but merely a partial oracle that is able to determine the outcome for each test case on each revision. 
This is an important distinction, because we do not
necessarily require a full specification of the system, 
but merely an answer to the question: `should test $t$ have passed or failed on revision $R$?'

\begin{dfn}[Oracle]
Given a test $t$ and a revision $R$, Oracle $\oracle{R}{t}$ gives the expected result for test $t$ in $R$.
\[
\begin{array}{c}
\mbox{$t$ is a {\bf true positive} for $R$ if $\fails{R}{t} \wedge \oracle{R}{t} = {\tt fail}$.}\\
\mbox{$t$ is a {\bf false positive} for $R$ if $\fails{R}{t} \wedge \oracle{R}{t} = {\tt pass}$.}\\
\mbox{$t$ is a {\bf true negative} for $R$ if $\passes{R}{t} \wedge \oracle{R}{t} = {\tt pass}$.}\\
\mbox{$t$ is a {\bf false negative} for $R$ if $\passes{R}{t} \wedge \oracle{R}{t} = {\tt fail}$.}\\
\end{array}
\]
\end{dfn}

We leave the question of how one determines the oracle deliberately unspecified.
In much of the literature on oracles~\cite{ebetal:oracle} there is an implicit assumption that there
exists a {\em single} ground-truth oracle for a software system, albeit one which may be unknown or only partially specified, but this is an unreasonable assumption.
This was pointed out as early as 1979 with the publication of a seminal paper  ~\cite{demillo:social}.
That paper was deemed,  by some, to be controversial~\cite{dijkstra:pamphlet} at the time.
Nevertheless, it has become increasingly apposite as the nature of Software Engineering has evolved over the intervening years~\cite{poh:incorrectness,hoare:how,hoare:icse96}.
Specifically, Software Engineering is a partly social process, 
so one engineer's (and/or one user's) oracle may differ from another.

We do not wish to enter into this detail here so we assume that, given a revision and a test, there is agreement on the oracle for the expected behaviour of the test on that revision.
For example, for practical purposes, we may consider that the decision is confined to the author of the pull request, since this is the person (or machine) who, for all practical purposes, will play the role of deciding whether a test signal is acceptable.

\subsection{Hardening Tests}
In this section, we formally define the concept of a hardening test: one that is designed to pass on the current version of the system and catch future regressions.

This is the kind of test produced by many test generation systems and approaches, including
Symbolic Execution systems~\cite{cadar:three-decades}, such as 
Klee~\cite{cadar:klee}, 
CUTE~\cite{sen:cute} and 
DART~\cite{godefroid:dart},
Search Based Systems such as 
EvoStuite~\cite{fraser:evosuite}
and
Austin~\cite{kletal:austin-ist},
Mutation-based test generation Systems such as
Javalanche~\cite{schuler:javalanche}
and
SHOM~\cite{mhetal:shom},
Replay Systems such as 
Fausta~\cite{kmetal:fausta}
and
JRapture~\cite{steven:jrapture}, 
Observation-based systems such as 
Carving ~\cite{elbaum:carving},
TestGen~\cite{mhetal:TestGen-obs}
and
Pankti~\cite{tiwari:mimicking}
and LLM-based systems such as
TestGen-LLM~\cite{mhetal:TestGen-LLM},
CoverUp ~\cite{pizzorno:coverup},
HITS~\cite{wang:hits}
and
ACH ~\cite{foster:mutation}.

Although these systems have different notions of coverage, such as mutants, serialised object observations and code, 
they all share the same approach to the oracle, which is that the behaviour of future changes should not disrupt the previously observed behaviour. 
This is the `regression oracle'~\cite{alshahwan:software}, in which a previous revision of the system acts as the ground truth oracle for future revisions.

A test $t$ for the revision $R$, is a {\em weak hardening test}, denoted $\hardening{R}{t}$,
if and only if it passes on $R$ and there exists a possible next revision on which it fails when it should.
More formally, Definition~\ref{dfn:weak-hardening} defines (weak) hardening.

\begin{dfn}[(Weak) Hardening Test]
\label{dfn:weak-hardening}
$\hardening{R}{t} \Leftrightarrow \passes{R}{t} \wedge \exists R'. \parent{R'} = R \wedge \fails{R'}{t} \wedge \oracle{R'}{t} = {\tt fail}$. 
\end{dfn}

Definition~\ref{dfn:weak-hardening} is `weak' in the sense that it does not require that the test, $t$ should correctly  pass on $R$, merely that it  should correctly fail on some future revision, $R'$.
The test `hardens' the code base against at least one buggy revision it might encounter in future.

A test $t$ for revision $R$, is a  {\em strong hardening} test denoted $\Shardening{R}{t}$, 
if it is a weak hardening test for $R$ and it passes correctly (true negative) for $R$.  
More formally, Definition~\ref{dfn:strong-hardening} defines strong hardening.

\begin{dfn}[(Strong) Hardening Test]
\label{dfn:strong-hardening}
\[
\begin{array}{c}
\Shardening{R}{t} \Leftrightarrow \hardening{R}{t} \wedge \oracle{R}{t} = {\tt pass} \\
\Leftrightarrow \\
\passes{R}{t} \\
\wedge \\
\oracle{R}{t} = {\tt pass} \\
\wedge \\
\exists R'. \parent{R'} = R \\
\wedge \\ 
\fails{R'}{t}\\
\wedge \\
\oracle{R'}{t} = {\tt fail}
\end{array}
\]
\end{dfn}

A strong hardening test is one for which we know that the passing signal on the current revision is a correct passing signal.
In general, this relies on oracle knowledge which cannot always be known at test generation time; it may not even be well defined, since engineers may disagree, with the consequence there is no ground truth.
Nevertheless, if there is a known agreed oracle outcome for the test case,  then it would be possible to determine, at test generation time, whether a weak hardening test is strong.
Indeed, in many cases it is worth the additional (minor) friction of asking the engineer to check 
that the test is correct in passing on the current revision.
This human-in-the loop check ensures that a weak hardening test is also a 
strong hardening test, thereby allowing us to proceed with greater confidence.

This is what we have done with our deployment of TestGen-LLM ~\cite{mhetal:TestGen-LLM} and ACH~\cite{foster:mutation}, 
both of which generate tests using LLMs, and propose these to the engineer who checks that the behaviour of passing on the current version of the system is correct.
In the case of TestGen-LLM the engineer also has to check whether the test will likely catch future regressions.
This extra obligation places a further burden on the human engineer that is removed by ACH, 
since it automatically provides an example of a fault that is {\em guaranteed} to be caught by the new proposed test.
As a result, ACH tests accepted by the engineer are strong hardening tests; the engineer  confirms that the test correctly passes on the current revision, 
while the ACH technology automatically generates the assurance that it catches at least one bug (the mutant) not caught by any existing test.

A strong hardening test is also a weak hardening test, by definition, but not vice versa.
When we refer to a `hardening test' (without qualification) we mean a weak hardening test (that is, one that may also happen to be strong).
When we wish to refer to a {\em weak} hardening test that is also {\em not a strong} hardening test, we call it a `{\em strictly} weak hardening' test.

Notice that a hardening test may be a coverage-improving test for some structural coverage criterion, 
such as statement coverage, but not necessarily so. 
The test may cover exactly the same code as already covered, yet do so in a different way to any existing test.

A hardening test distinguishes the behaviour of the current version of the system from that of some future change. 
In the terminology of mutation testing, where we think of a buggy $R'$ as a mutant of $R$, a hardening test {\em kills} the mutant.
It was this desire to go beyond coverage to automatically generate hardening tests that led to our recent work on mutation-guided LLM-based test generation~\cite{foster:mutation}.
ACH deployment guarantees, not only that tests are strong hardening, 
but also that they can catch  at least one bug that is uncaught by any existing test. 

\noindent
{\bf Running example:}
Throughout the following definitions we will present illustrative examples of tests and pull requests, 
based on a hypothesised simple
database application.
This illustrative application  involves a `product' object that includes a list of supplier locations.
This list should be initially empty.
It should be a list type because there could be more than one supplier, 
and it should be empty initially because we do not initially have any suppliers when we create the object.
Example~\ref{ex:weak-hardening} gives a unit test that we might write for this database application.

\begin{ex}[Weak Hardening Test]
\label{ex:weak-hardening}
\mbox{}\\
\vspace{-1em}
\scriptsize
\begin{verbatim}
    def weak_check_initial_supplier_locations():
      product = Product()
      locs = product.get_supplier_locs()
      assertTrue (not locs)
\end{verbatim}
\end{ex}

If a subsequent pull request were to introduce a bug that inadvertently assigns a nonempty list to the locations field on initialisation, Example~\ref{ex:weak-hardening} would catch that bug.
However, suppose the constructor {\tt Product()} {\em already} contains a bug (in the current revision) 
that allows the supplier locations field to be null (i.e., {\tt None} type in Python).
In this situation, Example~\ref{ex:weak-hardening} is incorrectly 
passing on the current revision, due to the familiar problem in Python that the unary {\tt not} operator returns 
{\tt True} when applied to the empty list, but also when applied to the null value  {\tt None}.

If the current revision contains this bug, 
then Example~\ref{ex:weak-hardening} is a {\em strictly weak} hardening test (i.e., it is not strong hardening).
We can create a better test, Example~\ref{ex:strong-hardening} below, which is strong hardening.
Example~\ref{ex:strong-hardening} fixes the assertion, replacing 
{\tt assertTrue (not locs)} 
with  
{\tt assertTrue (locs is not None and len(locs) == 0)}.

\begin{ex}[Strong Hardening Test]
\label{ex:strong-hardening}
\mbox{}\\
\vspace{-1em}
\scriptsize
\begin{verbatim}
    def strong_check_initial_supplier_locations():
      product = Product()
      locs = product.get_supplier_locs()
      assertTrue (locs is not None and len(locs) == 0)
\end{verbatim}
\end{ex}

Strictly weak hardening tests are problematic because they introduce a false sense of security; the current revision is wrongly believed to be more correct than it is. 
Furthermore, the same test weakness will extend to testing future possible revisions that may contain the same bug, 
and which the strictly weak test would allow to slip through as an {\em avoidable} false negative.
This is why we believe it is worth the minor friction  inherent in proposing newly generated  weak hardening tests to engineers 
to {\em review} (rather than simply automatically landing them into production without human review). 

The engineers' reviews ensure that we check the oracle and make the engineers 
the final arbiters of whether tests should land into production.
With this simple approach, 
we significantly increase the probability that the weak hardening tests we generate will prove to be {\em strong} hardening tests when deployed.

\subsection{Hardening Tests' Precision and Recall}
A test $t$ for revision $R$, is a  {\em Perfect Precision Hardening} test denoted $\PPhardening{R}{t}$, 
if it is a  hardening test for $R$ and, on any occasion when it fails, it is right to fail according to the oracle (no false positives).  
More formally, Definition~\ref{dfn:pp-hardening} defines a (Perfect Precision) Hardening Test.

\begin{dfn}[(Perfect Precision) Hardening Test]
\label{dfn:pp-hardening}
\[
\begin{array}{c}
\PPhardening{R}{t} \Leftrightarrow \hardening{R}{t} \\
\wedge\\ 
\forall R'. \fails{R'}{t} \Rightarrow \oracle{R'}{t} = {\tt fail} \\
\end{array}
\]
\end{dfn}

A test $t$ for revision $R$, is a  {\em Perfect Recall Hardening} test denoted $\PRhardening{R}{t}$, 
if it is a  hardening test for $R$ and, on any occasion when it should fail according to the oracle, it does fail (no false negatives).  
More formally, Definition~\ref{dfn:pr-hardening} defines a (Perfect Recall) Hardening Test.

\begin{dfn}[(Perfect Recall) Hardening Test]
\label{dfn:pr-hardening}

\[
\begin{array}{c}
\PRhardening{R}{t} \Leftrightarrow \hardening{R}{t} \\
\wedge\\ 
\oracle{R'}{t} = {\tt fail} \Rightarrow \forall R'. \fails{R'}{t} \\
\end{array}
\]
\end{dfn}

We can  conjoin definitions~\ref{dfn:pp-hardening} and \ref{dfn:pr-hardening} to produce a definition of a `perfect precision and recall hardening test'. 
A perfect precision and recall hardening test passes on the current revision, 
and correctly fails on every possible future buggy revision (and fails on no other revision).
That is, for every possible future revision, the test fails if and only if the revision is buggy with respect to the property tested.

A perfect precision and recall test is an idealised test that may be hard to achieve in practice.
Example~\ref{ex:perfect} illustrates how hard it is to achieve perfect precision and recall.
It defines a scenario in which a test may have {\em almost} perfect precision and recall.
Example~\ref{ex:perfect}, is a test that might, at first, appear to have perfect precision and recall.
It is certainly {\em hard} to construct an example where it incorrectly fails, or an example where it incorrectly passes, 
but not impossible.

\begin{ex}[Near perfect precision and recall test: {\tt check\_foo}]
\label{ex:perfect}
Suppose the function {\tt foo} has a single nullable formal parameter but its implementation assumes that the actual parameter passed is never null. 
For this example,  we  then write a unit test, {\tt check\_foo} that fails precisely when  a call site to {\tt foo} passes a null actual parameter.
\end{ex}

As the pull request in Example~\ref{ex:not-perfect} below shows, 
it is almost impossible for a test to attain perfect precision and recall.
Example~\ref{ex:not-perfect} does not challenge the belief that Example~\ref{ex:perfect} exhibits perfect recall, but it makes it clear that it does not exhibit perfect precision.
Once the pull request is submitted, 
it would become clear the {\tt check\_foo}'s failure when the call site passes null to {\tt foo}, 
is no longer a correct failure; it is a false positive.
Since the revision in Example~\ref{ex:not-perfect} exists (as a potential future revision), 
we must conclude that the precision of {\tt check\_foo} from  Example~\ref{ex:perfect} never {\em was} perfect.

\begin{ex}[A pull request that confirms that {check\_foo} does not have perfect precision]
\label{ex:not-perfect}
Consider a pull request for the function {\tt foo} 
that adds a check on the actual parameter, and gracefully behaves 
reasonably when that parameter should ever be null.
\end{ex}

\subsection{Measuring Tests' Precision and Recall}
\label{sec:measuringPR}
Although we cannot expect tests to have perfect precision and recall, we can {\em measure} the precision of a hardening test 
over a series of test outcomes.
When measuring precision we tend to use pseudo false positives~\cite{mhetal:ssbse18-keynote} as our criterion, 
because when an engineer rejects a test signal, even if it is a `correct' signal, then this represents the degree of friction, and that should be regarded as a kind of false positive; a pseudo false positive.
The distinction between pseudo-false positive and false positive is an interesting illustration of the difference between practical software testing and theoretical software testing.

We can measure precision, but tests' recall cannot be directly measured, because it relies on perfect knowledge about the ground-truth expected behaviour. 
Instead, we can {\em estimate} the recall using, for example, mutation testing (to sample the space of potential faults) or leak through of faults into production (to determine a lower bound on the number of false negatives).

Practicing software engineers  are typically forced to care more about precision than recall.
Perfect recall is unachievable, because perfect (i.e., exhaustive) testing is unachievable, and so recall is an aspiration.
Perfect precision is also unachievable, because the underlying questions are fundamentally undecidable in general.
Although we cannot hope achieve {\em perfect} precision and recall, 
{\em unnecessarily} low precision can lead to the entire automated testing technology being abandoned due to friction on the engineers who waste their time considering false positives.
By contrast, unnecessarily low recall is less risky to deployment.
In this section, we want to capture this inherent prioritization, 
and to suggest a different way of measuring precision and recall.

It is important to avoid  measuring and goaling on precision and recall independently.
It is also important to avoid aggregates such as the F1 score
that implicitly treat the two measurements as convertible commodities.
That is, by aggregating, we implicitly accept that it is possible to trade a certain amount of precision degradation for a certain amount of recall improvement.
This does not carry over well to the use case of industrial deployment of automated test generation.

Rather, we propose to goal on the $R@P=p$.
That is, the recall achievable when the precision ($P$) is fixed  to be no lower than $p$.
For example, $R@P=0.8$ is the recall achievable when we demand a precision of at least 0.8, thereby ensuring that the ratio of false positives to true positives is no worse than 1:4.

Choosing $0.8$ for $p$ is a conservative choice. 
We can be fairly sure that, 
for most applications, this will be a sufficiently high precision: 
engineers will be prepared to tolerate friction 20\% of the time, 
if the other 80\% of cases throw up true positive catches.




\subsection{Catching Tests}
In this section, we define the notion of a catching test: one that fails on the current revision and may thereby reveal a bug.

A  {\em weak catching} test $t$, for revision $R$, denoted $\catching{R}{t}$, is simply a test that fails on $R$.
More formally, Definition~\ref{dfn:weak-catching} defines a (Weak) Catching Test.

\begin{dfn}[(Weak) Catching Test]
\label{dfn:weak-catching}
\[
\begin{array}{c}
\catching{R}{t} \Leftrightarrow \fails{R}{t} 
\end{array}
\]
\end{dfn}

A weak catching test on some revision $R$ may give a false positive signal, so we define the concept of a strong catching test (that gives a true positive signal).
A  {\em strong catching} test $t$, for revision $R$, denoted $\Scatching{R}{t}$, is a weak catching test for which the failure on $R$ is regarded as a true positive according to the oracle.
More formally, Definition~\ref{dfn:strong-catching} defines a (Strong) Catching Test.

\begin{dfn}[(Strong) Catching Test]
\label{dfn:strong-catching}
\[
\begin{array}{c}
\Scatching{R}{t} \Leftrightarrow \fails{R}{t} \wedge  \oracle{R}{t} = {\tt fail}
\end{array}
\]

\end{dfn}

Example~\ref{ex:strong-catching} is  a pull request for which the hardening test from Example~\ref{ex:strong-hardening} would subsquently also become a (strong) catching test.

\begin{ex}[Strong Catching Test]
\label{ex:strong-catching}
Suppose a pull request inadvertently updates the code for the {\tt Product} constructor to  set the supplier location to {\tt None}.
\end{ex}

The test defined in Example~\ref{ex:strong-hardening} will correctly fail on the pull request in Example~\ref{ex:strong-catching}, 
so we conclude that the test is a  hardening and (strong) catching test for the pull request in Example~\ref{ex:strong-catching}.

As with weak and strong hardening tests, we refer to a weak catching test that is also not a strong catching test as a `{\em strictly} weak catching' test.

In general, Search Based Software Testing (SBST)~\cite{mh:icst15-keynote,mcminn:survey}  
systems such as Austin~\cite{kletal:austin-ist} for C
and EvoSuite~\cite{fraser:evosuite} for Java, 
and test generation systems based on concolic execution~\cite{godefroid:dart,sen:cute,cadar:three-decades} 
such as Klee~\cite{cadar:icse11} 
generate hardening tests.
They seek to catch future regressions, but use the current version as the oracle, so cannot catch new bugs, just regressions. 

By contrast, 
automated fuzzing techniques~\cite{manes:fuzzing}, 
such as AFL~\cite{zalewski:afl} and automated end-to-end test generation systems, 
such as 
Sapienz~\cite{mao:sapienz:16,mhetal:ssbse18-keynote},
WW~\cite{jaetal:ease21-keynote} 
Fausta ~\cite{kmetal:fausta}, 
and Google's RecSim~\cite{Ie:RecSym} 
can potentially find Catching JiTTests.
However, they use the implicit oracle (not the regression oracle) 
to determine whether the test has failed.
These Catching JiTTests can therefore  catch, only  those truly egregious and obvious faults 
that are flagged by the implicit oracle (such as memory violations and crashes).

\subsection{Regression Catching Test}
It could be that a catching test for $R$ is also a hardening test for the parent of $R$.
We call this a `regression catching' test, because a hardening test is a regression test;
the current behaviour of the system acts as an oracle (the `regression oracle'~\cite{alshahwan:software}).

A  test $t$ for revision $R$ is a {\em Weak Regression Catching Test} for $R$, $\Rcatching{R}{t}$,  if it is a weak catching test for $R$ and is also a  hardening test for the parent of $R$.
More formally, Definition~\ref{dfn:weak-regression-catching} defines a Weak Regression Catching Test.

\begin{dfn}[Weak Regression Catching Test]
\label{dfn:weak-regression-catching}
\[
\begin{array}{c}
\Rcatching{R}{t} \\
\Leftrightarrow \\
\passes{\parent{R}}{t} \\
\wedge \\
(\exists R'. \parent{R} = \parent{R} \wedge \fails{R'}{t} \wedge \oracle{R'}{t} = {\tt fail})\\
\wedge \\
\fails{R}{t}\\ 
\end{array}
\]
\end{dfn}

A  test $t$ for revision $R$ is a {\em Strong Regression Catching Test} for $R$, $\SRcatching{R}{t}$,  if it is a strong catching test for $R$ and is also a  hardening test for the parent of $R$.
More formally, Definition~\ref{dfn:strong-regression-catching} defines a Strong Regression Catching Test.

\begin{dfn}[Strong Regression Catching Test]
\label{dfn:strong-regression-catching}

\[
\begin{array}{c}
\SRcatching{R}{t} \\
\Leftrightarrow\\
\passes{\parent{R}}{t} \\
\wedge\\ 
(\exists R'. \parent{R} = \parent{R} \wedge \fails{R'}{t} \wedge \oracle{R'}{t} = {\tt fail}) \\
\wedge\\
\fails{R}{t}\\
\wedge \\
\oracle{R}{t} = {\tt fail}\\
\end{array}
\]

This simplifies to 

\[
\SRcatching{R}{t} \\
\Leftrightarrow\\
\passes{\parent{R}}{t} \wedge \fails{R}{t} \wedge \oracle{R}{t} = {\tt fail}
\]

\end{dfn}

Since the test defined in Example~\ref{ex:strong-hardening}  is both hardening and (strong) catching
for the pull requests in Example~\ref{ex:strong-catching}, we conclude that it is a (strong) regression catching test for this pull request.

\subsection{Functionality Catching Test}
A catching test does not necessarily need to be a regression catching test.
We want to push testing beyond the Regression Only Trap (ROT).
The revision for which the test catches a bug may have intended new functionality which the test catches.
If this is new functionality, then the catching test should not pass on the parent.
It may even be that the catching test, $t$ for a revision, $R$ does not even build on $\parent{R}$.
For example, suppose $R$ introduces a new method, $m$ and that $t$ tests the behaviour of $m$. 
Since $m$ does not exist in $\parent{R}$, then $t$ will not build on $\parent{R}$.

We  define a Functionality Catching test, as one that catches faults in $R$ but is not a hardening test for $\parent{R}$.
As with regression testing, it can be strong or weak depending on whether the failure on $R$ concurs with the oracle.

A  {\em Weak Functionality Catching} test $t$, for revision $R$ is a weak catching test for $R$ that is {\em not} a  hardening test for the parent of $R$.
More formally, Definition~\ref{dfn:weak-functionality-catching} defines a Weak Functionality Catching Test, $\Fcatching{R}{t}$.

\begin{dfn}[Weak Functionality Catching Test]
\label{dfn:weak-functionality-catching}
\[
\begin{array}{c}
\Fcatching{R}{t} \\
\Leftrightarrow\\
\neg(\hardening{\parent{R}}{t}) \wedge \catching{R}{t}     \\
\Leftrightarrow                                                \\
(\neg(\builds{\parent{R}}{t}) \vee \fails{\parent{R}}{t} \vee \\
\neg(\exists R'. \parent{R} = \parent{R} \wedge \fails{R'}{t} \wedge \oracle{R'}{t} = {\tt fail}) ) \\
\wedge                                                     \\
\fails{R}{t}                                               \\
\end{array}
\]

\end{dfn}

The definition of a weak functionality catching test is simple and intuitive: it is one that is catching but not hardening.
However, instantiating this simple definition, reveals several implicit criteria.
The test might not build on the parent, or it could build yet fail on the parent. 
There is also the possibility that there does not exist 
a revision that shares the same parent and for which the test correctly fails on that revision.
This last criterion arises due to the `weak' nature of the catch.

If we know that the test {\em correctly fails} on the current revision (a `strong' catch), then this awkward criterion disappears.
This can be seen in the definition of strong functionality catching test.
A  {\em Strong Functionality Catching} test $t$, for revision $R$ is a strong catching test for $R$ that is {\em not} a  hardening test for the parent of $R$.
More formally, Definition~\ref{dfn:strong-functionality-catching} defines a Strong Functionality Catching Test, $\SFcatching{R}{t}$.

\begin{dfn}[Strong Functionality Catching Test]
\label{dfn:strong-functionality-catching}
\[
\begin{array}{c}
\SFcatching{R}{t}\\
\Leftrightarrow\\
\neg(\hardening{\parent{R}}{t}) \wedge \Scatching{R}{t}    \\
\Leftrightarrow                                            \\
(\neg(\builds{\parent{R}}{t}) \vee \fails{\parent{R}}{t})  \\
\wedge                                                     \\
\fails{R}{t} \wedge \oracle{R}{t} = {\tt fail}             \\
\end{array}
\]
\end{dfn}

As an example, suppose a pull request updates the {\tt Product} object to introduce a method, {\tt distance}, 
that computes the distance to each supplier.
Suppose further that the pull request contains a bug that incorrectly omits an {\tt abs} operator that is required to ensure that differences in location are always positive.
Example~\ref{ex:strong-functionality-catching} is a strong functionality catching test.
That is, it will not build on the parent of the pull request (since it calls {\tt distance}, which does not exist in the parent).
It also fails correctly on the pull request, since the distance is incorrectly computed as a negative number.

\begin{ex}[Strong Functionality Catching Test]
\label{ex:strong-functionality-catching}
\mbox{}\\
\vspace{-1em}
\scriptsize
\begin{verbatim}
    def distance_test():
      test_supplier = MOCK_PRODUCT(Locations.LONDON)
      dist = test_supplier.distance()
      assertTrue(dist>=0)
\end{verbatim}
\end{ex}

Now, suppose that the author of the pull request fixes the bug as a result of the failure of the test in Example~\ref{ex:strong-functionality-catching}.
Now Example~\ref{ex:strong-functionality-catching} correctly passes on the fixed pull request. 
As such, the test now becomes a (strong) hardening test for the pull request revision, 
protecting it against future regressions in the {\tt distance} method.

\section{Eight  Categories  of Test}
\begin{figure}

\centerline{\includegraphics[width=1.0\linewidth]{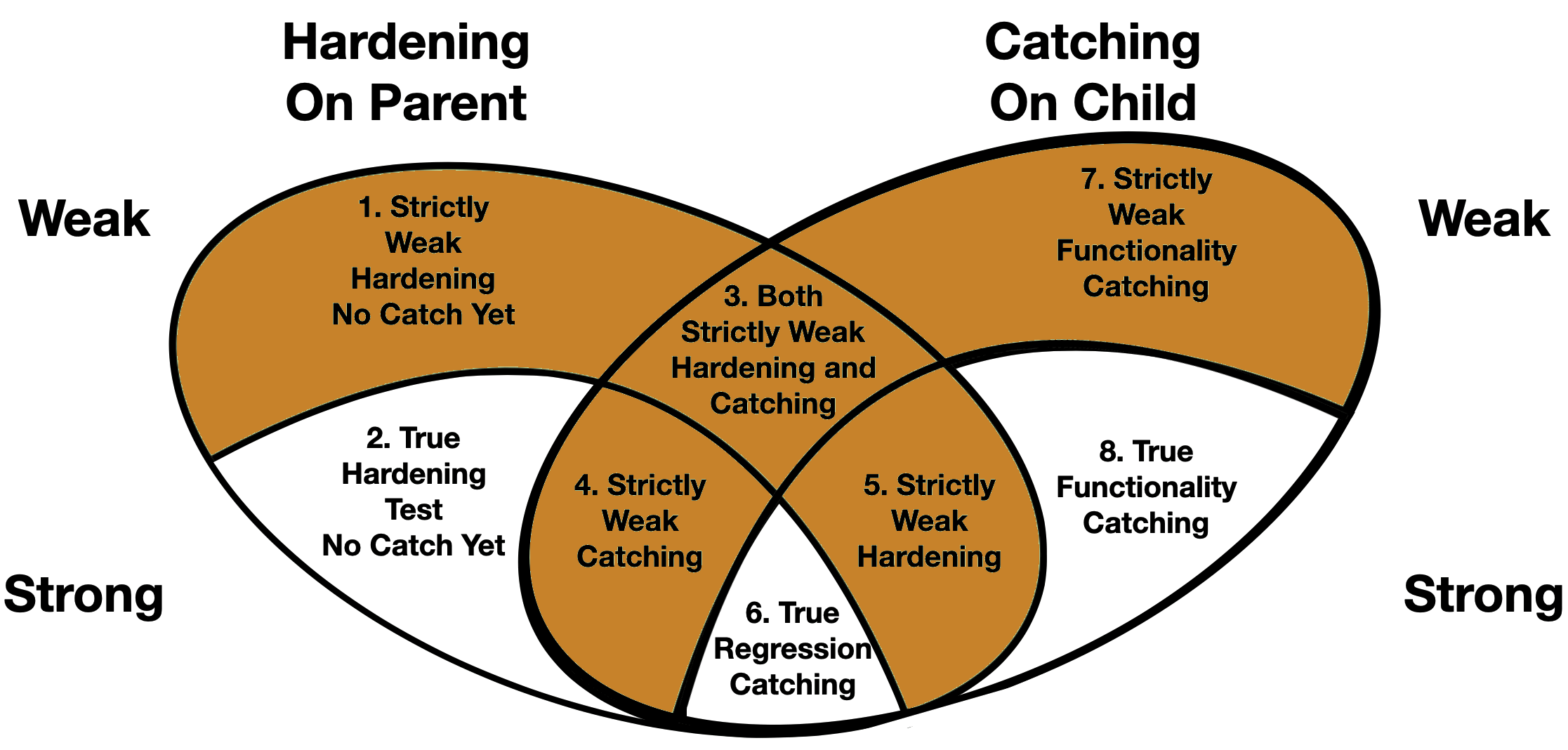}}
\caption{Hardening and catching Venn Diagram:  
Strong is a subset of weak. 
A test can be both (strong/weak) hardening on parent and also (strong/weak) catching on child. 
Each region of the Venn diagram corresponds to an interesting and important test behaviour.}
\label{fig:top_level}
\end{figure}

The 8 possible combinations of strong/weak hardening on parent  and catching on child and their intersections are shown in the Venn diagram in Figure~\ref{fig:top_level}.
In each of these eight regions there reside tests that exhibit interesting behaviours, 
five of which (the shaded regions) give rise to misleading test signals, 
while only three of which lead to reliable true signal.
In the remainder of this section we explain the eight categories of test depicted in Figure~\ref{fig:top_level}.

\noindent
{\bf Five misleading classes of signal (shaded regions in Figure~\ref{fig:top_level})}: 
A test that is strictly weak hardening but not catching (Region 1)   suggests that all is well, when it is not. 
A test that is strictly weak catching but not hardening (Region 7) may appear to catch new functionality 
but, because it is strictly weak catching, it is a `fake'  functionality catching test.
The remaining three misleading signals include two combinations of strictly weak and strong hardening/catching (Regions 4 and 5), and finally, the most misleading of all (Region 3): both strictly weak hardening and strictly weak catching.
A test that resides in this most misleading category might fail to catch a bug and subsequently `complain' (i.e., incorrectly fail) when it is fixed.

\noindent
{\bf Three classes of true signal}:
A test that is strong hardening, but not catching (Region 2), 
is simply a good hardening test that has yet to catch anything.
A test that is strong catching but not hardening (Region 8)
is a true functionality catching test.
A strong catching test that is also a strong hardening (Region 6) is a true regression catching test.

\noindent
{\bf Research Challenge 1}: find techniques that can help reduce the risk of miss-classification of tests according to this classification.

\section{Deployment of Hardening and Catching Tests }
\label{sec:case-by-case}

In this section we review the options for deploying hardening and catching tests. 
In the case of timely hardening, this provides a  way to formally analyse  
the (current widely-adopted) 
approach used in research and practice for automated regression test generation.
For  JiTTests, the analysis reveals the complexities and trade-offs in the decision processes, leading to an initial set of proposed deployment options.
We do not claim these proposed deployment options are unequivocal nor that they are comprehensive but rather, 
that they illustrate how the formal analysis together with a case-by-case reasoning can shed light on the trade-offs involved.

\subsection{Cases of Behaviour for Timely Hardening Tests}
\label{sec:deploying-hardening}
Table~\ref{tab:truth-table-hardening} captures\footnote{The tables in this paper use a traffic light colour-coding scheme which is best viewed in colour, rather than in black-and-white. Green indicates a positive test experience, in which there is no friction from misleading signal and there is test benefit. Red denotes friction: the test does not catch bugs but rather slows down the engineering development. Amber indicates possible friction, that is deemed to be probably tolerable.} 
the possible combinations of outcomes of a test generated on a revision, which we call the parent, 
and that seeks to generate a hardening timely test for this revision.
Such a test seeks to harden the parent against future regressions in a subsequent revision, which we call the child.
In this mode of deployment, we generate tests on the parent pull request, 
where the test passes on this parent, and we have some unknown future children, 
for which we want to protect the code.
From this case-by-case analysis, we can see that any test that passes on the current version of the system is
worthy of consideration.

In order to distinguish between strictly weak and strong hardening tests, 
we can use the simple expedient of reporting the generated test to the engineer.
Indeed, this is what we currently do with the ACH hardening tool ~\cite{foster:mutation}.
It is also what many other other test generation tools do.
Reporting the test to the engineer is not necessary: one could simply auto-land all passing generated tests into production, 
but such auto landing would be risky.  
It risks landing strictly weak hardening tests, 
or even tests that are {\em not} hardening at all because they can catch no future regression.

\noindent 
{\bf Actionable Conclusion}: 
By seeking the engineer's review, we eliminate three (of the five) problematic categories (regions 1, 3 and 5 deplicted  in Figure~\ref{fig:top_level}), thereby dramatically  reducing potential future confusion from `fake' test signals.
Furthermore, the effort  required from the engineer is low; they merely need to determine whether the assertion is reasonable, 
and therefore the test is right to pass on the current revision. 
This is a good example the principal that

\begin{quote}
``A relatively small amount of  timely oracle information can dramatically enhance test effectiveness.'' 
\end{quote} 

In this case, a check that may take an engineer seconds to perform 
removes three fifths of the problematic test categories.
Rather than thinking of the oracle as some unknowable theoretical construct, there is a more `engineering orientated viewpoint': 
The oracle is not a discrete logical construct, but better thought of as a continuous  resource on which we can draw.
Since it is an {\em expensive} resource on which to draw, we need to make the engineering trade-off that determines when the cost is worth the benefit.
The only situation where auto-landing generated tests might be safe from these problematic categories, is when tests are disposable. 
For example, where they are generated to detect any changed behaviour after some a specific meaning-preserving refactoring, and subsequently discarded.

\begin{table*}
\caption{The possible outcomes for generated timely tests.  
The  test is  generated for the current revision (the parent). 
We distinguish the possible outcomes for the execution on some future child of the current revision. 
In describing the outcome for a child  we are also describing the outcome for some future `child of a child of a ...'  
for some arbitrary number of  revisions for which the test passes.
A test that fails on the child is one of the catching cases from Figure~\ref{fig:top_level}, whereas a test that passes on both  the parent and child transitions from one hardening case (for parent) to another (for the child), hence the `$\rightarrow$' in column 6.
}
\label{tab:truth-table-hardening}
\begin{tabularx}{\textwidth}{|p{1.1cm}|p{1.1cm}|p{0.8cm}|p{0.8cm}|p{1.8cm}|p{0.9cm}|X|}
 \hline 
\multicolumn{2}{|c|}{outcome  on}           & \multicolumn{2}{|c|}{oracle  for} &  type              & In                       &                                                                    \\
parent  & child                             & parent     &  child               &  for child         & Figure~\ref{fig:top_level} & Consequences if we were to accept the generated test's signal      \\
\hline 
\hline
\multicolumn{7}{|c|}{{\bf When tests were landed to harden the parent and now they are executed on some future child ...} }\\
\hline
{\tt pass}             & {\tt pass}         & {\tt pass}       & {\tt pass}     &  true negative   & $2 \rightarrow 2$ & \textcolor{green} {Strong Hardening test passes: All is well; move fast} \\
{\tt pass}             & {\tt pass}         & {\tt fail}       & {\tt pass}     &  true negative   & $1 \rightarrow 2$ & \textcolor{green} {Strictly Weak Hardening test passes: the child fixes an uncaught bug} \\
{\tt pass}             & {\tt fail}         & {\tt pass}       & {\tt fail}     &  true positive   & \multicolumn{1}{|c|}{6} & True regression catching: \textcolor{green}{Catches the child's regression}\\
{\tt pass}             & {\tt fail}         & {\tt fail}       & {\tt fail}     &  true positive   & \multicolumn{1}{|c|}{5} & Aparent regression is more likely to be a \textcolor{green} {Failed error propagation~\cite{kaetal:analysis}}\\
{\tt pass}             & {\tt pass}         & {\tt fail}       & {\tt fail}     &  false negative  & $1 \rightarrow 1$ & \textcolor{orange}{Strictly Weak Hardening test continues to miss a bug}\\
{\tt pass}             & {\tt pass}         & {\tt pass}       & {\tt fail}     &  false negative  & $2 \rightarrow 1$ & \textcolor{orange}{Strong Hardening test misses a bug introduced by the child}\\
{\tt pass}             & {\tt fail}         & {\tt pass}       & {\tt pass}     &  false positive  & \multicolumn{1}{|c|}{4} & \textcolor{red}   {Strong Hardening test wrongly blocks the child}\\
{\tt pass}             & {\tt fail}         & {\tt fail}       & {\tt pass}     &  false positive  & \multicolumn{1}{|c|}{3} & \textcolor{red}   {Strictly Weak Hardening test wrongly flagged a fix  as a new failure}\\
\hline
\hline
\multicolumn{7}{|c|}{{\bf When tests were landed to harden the parent, but some future child makes them obsolete ... }}\\
\hline
{\tt pass}               & no build         & {\tt pass}        & ---      &  ---   & $2 \rightarrow \bot$ & Strong Hardening test retired: \textcolor{green} {child removes tested functionality} \\  
{\tt pass}               & no build         & {\tt fail}        & ---      &  ---   & $1 \rightarrow \bot$ & Strictly Weak Hardening test retired: \textcolor{green} {Test was weak anyway}\\  
\hline
\end{tabularx}
\end{table*}

Overall, therefore, we think it is safe to deploy hardening test generation and to land into production any tests that pass 
consistently (i.e., non flakily~\cite{mhpoh:scam18-keynote,mcetal:flakime,luo:flaky}) 
on the current revision, that are acceptable to human reviewers.
While this can have problems, as indicated by the four problematic rows attributed to false negatives and false positives in Table~\ref{tab:truth-table-hardening}, 
any such risks are merely specific instances of the risks that already pertain to human-written tests.

\noindent
{\bf Deploying  Hardening Timely Tests}:
Based on Table~\ref{tab:truth-table-hardening}, when we generate a hardening test from a current revision, and the test passes, 
we will offer the test up for review by an engineer who may choose to land it into production.
At the heart of this decision, resides  the {\em core belief in the value of testing itself}:

\begin{quote}
``The risk of future false positives is outweighed by the chance of catching future real bugs.''
\end{quote}


\noindent
{\bf Research Challenge 2: automatically distinguish weak from strong hardening}: find techniques that can take partial oracle information  
and use this to improve the chance that a weak hardening test is strong, to further reduce human oracle effort.

\noindent
{\bf Research Challenge 3: automatically distinguish regression catching from false positive}:
find techniques that reduce the risk that a test will raise false positive signal, due to strictly weak catching.

\subsection{Cases of Behaviour for JiTTests}
We are specifically interested in the case of generated JiTTests, because a JiTTest is generated at the last occasion in which it is possible to catch a bug before it lands into production. 
This makes this particular testing paradigm  highly important and impactful.
Despite this importance, there has been very little discussion and analysis of the opportunities and challenges for generation of such JiTTests.
Table~\ref{tab:truth-table-catching} collects the possible outcomes for  attempts to generate a
JiTTest  for a revision $R$, and the consequences of accepting the signal from the generated test.

\begin{table*}
\caption{The possible outcomes for  generation of  Just-in-Time Tests (JiTTests). 
In this mode of deployment, tests are generated on the fly when a revision is first submitted. 
A test is generated when a revision $R$ is submitted, based on the revision $R$,  and its parent, $\parent{R}$.  
The outcome of test execution on both the revision and the parent can be taken into account when determining whether to land the test, report its signal (but not land the test), or to  discard it completely.
In Column 6, JiTTests that fail on $R$ correspond to one of the categoiries from Figure~\ref{fig:top_level}, for which a decision must be made about whther to give the signal to the engineer.  
Cases that do not fail on $R$  may correspond to transitions between categories ($\parent{R} \rightarrow R$). 
}

\label{tab:truth-table-catching}
\begin{tabularx}{\textwidth}{|p{1.2cm}|p{1.1cm}|p{1.2cm}|p{1.1cm}|p{1.8cm}|p{0.9cm}|X|}
 \hline 
\multicolumn{2}{|c|}{outcome  on}           & \multicolumn{2}{|c|}{oracle  for} &  type            & In  &             \\
$\parent{R}$  & $R$                         & $\parent{R}$     &  $R$           &  for $R$         & Figure~\ref{fig:top_level} & Consequences of landing the JiTTest and reporting its signal\\
\hline 
\hline

\multicolumn{7}{|c|}{{\bf Category 1 (Parent Hardening): tests that pass on the parent  ...} }\\
\hline
{\tt pass}             & {\tt pass}         & {\tt pass}       & {\tt pass}     &  true negative   & $2 \rightarrow 2$        &\textcolor{green} {Strong Hardening JiTTest hardens $R$} \\
{\tt pass}             & {\tt pass}         & {\tt fail}       & {\tt pass}     &  true negative   & $1 \rightarrow 2$        &\textcolor{green} {Strictly Weak Hardening on parent (missed bug likely fixed)} \\
{\tt pass}             & {\tt fail}         & {\tt pass}       & {\tt fail}     &  true positive   &  \multicolumn{1}{|c|}{6} &\textcolor{green} {True Regression catching JiTTest}\\
{\tt pass}             & {\tt fail}         & {\tt fail}       & {\tt fail}     &  true positive   &  \multicolumn{1}{|c|}{5} &\textcolor{green} {Strictly Weak Hardening JiTTest}\\
{\tt pass}             & {\tt pass}         & {\tt fail}       & {\tt fail}     &  false negative  & $1 \rightarrow 1$        &\textcolor{orange}{Strictly Weak Hardening JiTTest misses a long-standing issue}\\
{\tt pass}             & {\tt pass}         & {\tt pass}       & {\tt fail}     &  false negative  & $2 \rightarrow 1$        &\textcolor{orange}{Strong Hardening JiTTest missed a bug introduced by $R$}\\
{\tt pass}             & {\tt fail}         & {\tt pass}       & {\tt pass}     &  false positive  &  \multicolumn{1}{|c|}{4} &\textcolor{red}   {Strictly Weak Catching JiTTest wrongly blocks change}\\
{\tt pass}             & {\tt fail}         & {\tt fail}       & {\tt pass}     &  false positive  &  \multicolumn{1}{|c|}{3} &\textcolor{red}   {Doubly Weak JiTTest wrongly blocks a fix}\\
\hline
\hline
\multicolumn{7}{|c|}{{\bf Category 2 (Child Hardening): tests that fail on the parent and pass on the child  ...} }\\
\hline
{\tt fail}             & {\tt pass}         & {\tt pass}       & {\tt pass}     &  true negative   & $7 \rightarrow 2$ &\textcolor{green} {Strong Hardening JiTTest: $R$ fixes a broken test} \\
{\tt fail}             & {\tt pass}         & {\tt fail}       & {\tt pass}     &  true negative   & $8 \rightarrow 2$ &\textcolor{green} {Strong Hardening JiTTest correctly identifies a fix}\\
{\tt fail}             & {\tt pass}         & {\tt pass}       & {\tt fail}     &  false negative  & $7 \rightarrow 1$ &\textcolor{orange}{Strictly Weak Hardening JiTTest lets through new bug} \\
{\tt fail}             & {\tt pass}         & {\tt fail}       & {\tt fail}     &  false negative  & $8 \rightarrow 1$ &\textcolor{orange}{Strictly Weak Hardening JiTTest lets an existing bug remain}\\
\hline
\hline
\multicolumn{7}{|c|}{{\bf Category 3 (New Functionality Tests): tests that fail to build on the parent ... }}\\
\hline
no build               & {\tt pass}         & ---       & {\tt pass}     &  true negative   & $\bot \rightarrow 2$        &\textcolor{green} {Strong Hardening for new functionality introduced by $R$} \\  
no build               & {\tt fail}         & ---       & {\tt fail}     &  true positive   &  \multicolumn{1}{|c|}{8} &\textcolor{green} {Strong Hardening/Catches $R$'s new functionality bug} \\  
no build               & {\tt pass}         & ---       & {\tt fail}     &  false negative  & $\bot \rightarrow 1$        &\textcolor{orange}{Strictly Weak Hardening bakes in incorrect new functionality}\\  
no build               & {\tt fail}         & ---       & {\tt pass}     &  false positive  &  \multicolumn{1}{|c|}{7} &\textcolor{red}   {Strictly Weak Catching blocks a perfectly good change}\\  
\hline
\hline
\multicolumn{7}{|c|}{{\bf Category 4 (Cannot Land): tests that fail to build on  the child  ... }}\\
\hline
\multicolumn{2}{|c|}{outcome  on}           & \multicolumn{2}{|c|}{oracle  for} &  type            &  &           \\
$\parent{R}$  & $R$                         & $\parent{R}$     &  $R$           &  for $R$         &  & Consequences of reporting the JiTTest signal to the engineer \\
\hline
{\tt pass}               & no build         & {\tt pass}       & ---            &  ---             &  $2 \rightarrow \bot$   & \textcolor{green} {Strong Hardening but tested functionality is removed by $R$} \\  
{\tt fail}               & no build         & {\tt fail}       & ---            &  ---             &  $8 \rightarrow \bot$   & \textcolor{green} {Strong Catching JiTTest may flag the bug removed by $R$} \\  
{\tt fail}               & no build         & {\tt pass}       & ---            &  ---             &  $7 \rightarrow \bot$   & \textcolor{orange}{Strictly Weak Catching false positive, but it is removed }\\  
{\tt pass}               & no build         & {\tt fail}       & ---            &  ---             &  $1 \rightarrow \bot$   & \textcolor{orange}{Strictly Weak Hardening weakness means little valuable signal}\\  
\hline
\hline
\multicolumn{7}{|c|}{{\bf Category 5 (Likely Discard): tests that fail on both the parent and child ...} }\\
\hline
\multicolumn{2}{|c|}{outcome  on}           & \multicolumn{2}{|c|}{oracle  for} &  type            & &            \\
$\parent{R}$  & $R$                         & $\parent{R}$     &  $R$           &  for $R$         & & Consequences of discarding the JiTTest and ignoring its signal\\
\hline
{\tt fail}             & {\tt fail}         & {\tt pass}       & {\tt pass}     &  false positive  &  \multicolumn{1}{|c|}{7} & \textcolor{green} {Strictly Weak Catching JiTTest should be discarded}\\
{\tt fail}             & {\tt fail}         & {\tt fail}       & {\tt pass}     &  false positive  &  \multicolumn{1}{|c|}{7} & \textcolor{green} {Strictly Weak Catching so would have rejecting a correct fix}\\
{\tt fail}             & {\tt fail}         & {\tt pass}       & {\tt fail}     &  true positive   &  \multicolumn{1}{|c|}{8} & \textcolor{orange}{Discarding Strong Catching JiTTest meant missing a new bug}\\
{\tt fail}             & {\tt fail}         & {\tt fail}       & {\tt fail}     &  true positive   &  \multicolumn{1}{|c|}{8} & \textcolor{orange}{Discarding Strong Catching means missing existing bug } \\
\hline
\end{tabularx}
\end{table*}

For a  JiTTest, the test generation tool has available to it, both the pull request, $R$,  and its parent, $\parent{R}$, 
at test generation time.
The table groups the analysis results for the current revision and its parent into five categories, with a sub-table for each category.
In this section, we consider each category and turn.

\noindent
{\bf Category 1 (Parent Hardening): tests that pass on the parent}. 
A hardening test may have been generated some months or even years before the first time it fails in production on a new revision.
By contrast, a  JiTTest is generated on the fly when a revision is submitted. 
Where the test passes on the parent revision, 
the JiTTest may be a hardening test for the parent, 
although it is not timely for the parent, only for the child, since it is a JiTTest.

Where the test passes on the parent,
the case-by-case analysis is isomorphic to the case-by-case analysis for the  hardening timely test.
This is shown in the first eight rows of Table~\ref{tab:truth-table-catching}.
Although isomorphic, 
the consequences of false positives are more problematic than for  hardening timely tests, 
precisely because of the just-in-time nature of the test generation process.

Since the tests are generated just-in-time for the revision, 
it is likely that there is no existing test that fails.
If there were, then we would not need to generate any further tests.
Furthermore, there may simply be {\em  no} existing tests at all (either passing or failing).
This is precisely the scenario in which we would need to generate JiTTests, to either increase confidence in the revision under test or stop a bug form landing into production.
For the same reason, when a JiTTest fails on a revision, it may be the only test that provides any signal, whether passing or failing. 
Therefore, the risk of high friction from a newly created false positive is considerably higher than for a long-standing hardening timely test.

We may choose to suppress the signal of a  JiTTest to avoid undue friction.
The balance between risk and benefit will likely be determined differently by different organisations and different projects within an organisation.
Notwithstanding these organizational preferences, 
 a general principle applies to all: the risk of friction is higher for  JiTTests than for merely timely tests.

\noindent{\bf Actionable Conclusion}: 
As a result of this analysis, a failing JiTTest signal may be regarded as a `trigger' 
to do further testing or analysis before troubling the engineer with the signal.
By contrast passing JiTTest signal still plays the role of giving the engineer additional confidence, 
and such a passing JiTTest may also be a hardening JiTTest.
It therefore makes sense to have a deployment route for JiTTests, if only to create hardening 
JiTTests as the side effect of attempting to generate catching JiTTests.

\noindent
{\bf Category 2 (Child Hardening): tests that fail on the parent and pass on the child}. 
For all four cases in Category 2, the JiTTest fails on the parent but passes on the current revision.
In any such case, it is relatively safe to give the signal to the engineer.
A Category 2 test that is a hardening JiTTest is, 
by definition, both hardening and timely for $R$, 
although it is neither hardening nor timely for its parent.

When the test is a true negative for the current revision, its behaviour not only generates a new hardening test for the future (as a  hardening timely test would), it can yield additional signal value since it is just-in-time.
For example, the test may show that the developer's change has fixed (or improved) the behaviour of the parent revision.
Such fail $\rightarrow$ pass  JiTTests therefore add value over and above purely generating new hardening timely tests 
(which they also do `for free').
Also, fortunately, when the test is a false negative pass, 
the negative consequences are relatively low, as the case-by-case analysis reveals.
This observation mimics the analysis for  hardening timely tests in Section~\ref{sec:deploying-hardening}.

\noindent{\bf Actionable Conclusion}: 
For both Categories 1 and 2 the generated test is a hardening timely test for the child but not for the parent.
It is safe to give this timely hardening signal to the engineer when the test is generated for $R$.

\noindent
{\bf Category 3 (New Functionality Tests): tests that fail to build on the parent}. 
Where a JiTTest fails to build on the parent of the current revision and passes on the current revision, 
this is perfectly safe to land (as the case analysis in Table~\ref{tab:truth-table-catching} shows). 
Such a test will become a new hardening test which, like others, may `bake in' wrong functionality, but there is the hope that other tests will catch that.
Furthermore, since the test fails to build on the parent, it is likely that this test is testing precisely the new functionality added by the revision.
Once again, as with Category 2 (for tests that fail on the parent), 
the just-in-time nature of the test gives extra signal because it does not build on the parent.

\noindent{\bf Actionable Conclusion}: 
From the analysis of Categories 2 and 3, we conclude 
that it is {\em always} wise to run a generated test on the parent of a revision 
(to obtain this extra signal).
This is not something that current test tools tend to do, 
but our analysis suggests that it is not only worthwhile.
Test generation might {\em specifically target} the generation of such tests, 
and check/report test behavior on the parent of the tested revision.

\noindent
{\bf Category 4 (Cannot Land): tests that fail to build on  the child}. 
Where a JiTTest fails to build on the current revision, it clearly cannot be landed as a new test.
Nevertheless, its signal may be useful to the engineer. 
For example, if it previously failed on the parent, then it may indicate a fix or an improved functionality in the current revision.
The engineer may even choose to cite the generated test  as {\em documentation} for the improvement that their revision achieves.

\noindent{\bf Actionable Conclusion}: 
From the analysis of Category 4, 
we conclude that we cannot land tests that fail to build on the child, but it is always worth giving the signal to the engineer.

\noindent
{\bf Category 5 (Likely Discard): tests that fail on both the parent and child}. 
Finally, JiTTests that fail on both the parent and the current revision will typically have to be discarded.
The only corner case is one where we believe we have discovered a long-standing or new bug, but there will need to be an additional way to gain confidence that this is what the test is revealing rather than a false positive.

The risk of friction from a false positive is relatively high.
The chances of finding such a  bug with a test that fails on the parent and the current revision is low.
Therefore, it seems likely that the balance probabilities is in favour of discarding all such tests.

\noindent{\bf Actionable Conclusion}: 
From the analysis of Category 5, without new techniques, we may sadly have to discard tests that fail on both the parent and the child.
This is the least certain of our actionable conclusions from the analysis in this section. 
Futhure research will hopefully find ways to distinguish sub-categories; one aspect of the Catching JiTTest challenge (see Section~\ref{sec:challenge}).

\noindent
{\bf Deploying catching tests}:
We summarize the foregoing actionable conclusions to
document the eight possible outcomes that can be observed, irrespective of the oracle
in Table~\ref{tab:truth-table-catching-decisions}.

\begin{table*}
\caption{Suggested JiTTest deployment approach based on signal from parent and current revision.  Based on the case by case analysis in Table~\ref{tab:truth-table-catching} we give a proposed deployment approach for each case. 
Of course, unlike the analysis from Table~\ref{tab:truth-table-catching}, oracle information is assumed to be unavailable, and therefore we do not know whether signal is a true/false positive/negative.}
\label{tab:truth-table-catching-decisions}
\begin{tabularx}{\textwidth}{|p{1.2cm}|p{1.2cm}|p{1.6cm}||X|X|}
\hline
\multicolumn{5}{|c|}{{\bf Category 1: When JiTTest passes on parent of current revisions  ...} }\\
\hline
\multicolumn{2}{|c|}{outcome  on}               &            & &  \\
$\parent{R}$  & $R$                             &  Decision   & Possible undesirable consequences & Best case consequences \\
\hline
{\tt pass}             & {\tt pass}             & Land Test    & \textcolor{orange} {Bakes in long standing issue}                                 & \textcolor{green} {Catches future regressions}\\
{\tt pass}             & {\tt fail}             & Report Fail  & \textcolor{red}    {False positive}; \textcolor{orange}{but other tests may pass} & \textcolor{green} {Catches a current regression} \\
\hline
\hline
\multicolumn{5}{|c|}{{\bf Category 2: When JiTTest fails on parent but passes on current revision ...} }\\
\hline
\multicolumn{2}{|c|}{outcome  on}               &            & &  \\
$\parent{R}$  & $R$                             &  Decision & Worst case consequences of landing new test & Best case consequences of landing new test\\
\hline
{\tt fail}             & {\tt pass}             & Land Test & \textcolor{orange} {Bakes in newly created issue}& \textcolor{green} {Catches future regressions in  new functionality} \\
\hline
\hline
\multicolumn{5}{|c|}{{\bf Category 3: When  JiTTest does not build on the parent of the current revision ... }}\\
\hline
\multicolumn{2}{|c|}{outcome  on}               &         & &     \\
$\parent{R}$  & $R$                             &   Decision  & Worst case consequences & Best case consequences\\
\hline
no build               & {\tt pass}             & Land Test   & \textcolor{orange} {Bakes in newly created issue}                                & \textcolor{green} {Catches future regressions in  new functionality} \\  
no build               & {\tt fail}             & Report Fail & \textcolor{red} {False positive}; \textcolor{orange}{but other tests may pass}   & \textcolor{green} {Catches bug in new functionality} \\  
\hline
\hline
\multicolumn{5}{|c|}{{\bf Category 4: When JiTTest cannot land but may still give useful signal on the revision  ... }}\\
\hline
\multicolumn{2}{|c|}{outcome  on}              & &                                              &  \\
$\parent{R}$  & $R$                            & & Worst case consequences of reporting signal  & Best case consequences of reporting signal \\
\hline
{\tt pass}               & no build            & Give Signal  & \textcolor{orange} {Wrongly claims that parent was correct}   & \textcolor{green} {Confirms that behaviour was removed} \\  
{\tt fail}               & no build            & Give Signal  & \textcolor{orange} {Wrongly claims that parent was incorrect} & \textcolor{green} {Helps to document a fix}  \\  
\hline
\hline
\multicolumn{5}{|c|}{{\bf Category 5: When JiTTest fails on the current revision and its parent ...} }\\
\hline
\multicolumn{2}{|c|}{outcome  on}            & &                                                  &  \\
$\parent{R}$  & $R$                         & & Worst case consequences of discarding  JiTTest & Best case consequences of discarding  JiTTest\\
\hline
{\tt fail}             & {\tt fail}          & Discard Test & \textcolor{orange} {Loses a bug catch on current revision} & \textcolor{green} {Loses a potentially misleading test} \\
\hline
\end{tabularx}
\end{table*}


\section{Research Challenge 4: The Catching JiTTest Challenge}
\label{sec:challenge}

Earlier in the paper, we discussed Research Challenges 1 -- 3, which concerned classification and hardening tests.
These three challenges are important, 
but the most challenging of all 
(and likely the most impactful) 
is Research Challenge 4: The Catching JiTTest Challenge, 
which we summarise as follows:

\begin{quote}
    ``Given a buggy pull request and its parent, 
    automatically generate a test that can reveal  bugs in the pull request,
    with low risk of false positive failure on non-buggy pull requests. ''
\end{quote}

We call this the `Catching JiTTest Challenge'.
It is essentially to improve the value of JiTTest 
R@P=$p$ (See Section~\ref{sec:measuringPR}), 
for some suitably high choice of precision threshold $p$, such as 0.8.

Suppose a JiTTest, $t$, generated just-in-time for a revision, $R$, passes on $R$ and also happens to pass on $\parent{R}$. 
This test, $t$, is simply a hardening JiTTest (for $R$). 
However, if $t$ fails (or fails to build) on $\parent{R}$,  
then $t$ is a specific new kind of test case that can be found {\em only} using the Just-in-Time paradigm:
It could not be landed at any time before $R$ has landed, because it would land a failing test signal or even break the build.

The Catching JiTTest Challenge thus opens up a whole new class of test cases that can be generated.
In order to tackle the challenge, we are inherently faced with a particularly pernicious example of the Oracle Problem:
the only oracle information may be that in the pull request itself.
Such information can be exceptionally parsimonious.

\subsection{The Parsimonious Pull Request Problem and the Oracle Scavenging Solution}
As an extreme example of a parsimonious pull request, consider the following hypothesised (but entirely plausible) pull request:

\begin{tabbing}
summary: \= sp\=  TSIA  \kill \\
    {\bf title}: \> \> add location details to lookup query \\
    {\bf summary}: \> \> TSIA \\
    {\bf Test Plan}: \> \> YOLO \\
\end{tabbing}

Let us call this pull request $TSIA$.
The acronym `TSIA' stands for `Title Says It All', while the acronym `YOLO' stands for `You Only Live Once'.
Engineers may believe this level of detail is sufficient for the reviewer. 
Therefore, it is also the only information available to any automatic technology seeking to design tests.

\noindent
{\bf Code Summarization May Prove Insufficient}:
Of course, we could seek to use LLMs for code summarization~\cite{zhu2019automatic} to provide a better description in the summary.
There has been much research work on this problem~\cite{mhetal:LLM-survey}, and it would undoubtedly help the {\em human} reviewer.
Nevertheless, when  our goal is to  automatically provide a better test plan than `YOLO', 
we would need to be  cautious to ensure that our LLM-based test generation techniques do not become victims to the obvious circularity involved by also using LLMs to infer the code summaries. 

\noindent
{\bf Grounds for Optimism: Plenty of Computation Time}:
As revealed in Section~\ref{sec:online}, the timing constraints for online test generation are considerably more relaxed than might, at first, be thought. 
In deployment to continuous integration we are typically afforded approximately eight hours
to generate  JiTTests.
This allows for considerable static and dynamic analysis, 
and other automated techniques to be deployed to determine what signal, if any, should be given to the pull request reviewer.

\noindent
{\bf Grounds for Optimism: Tackling the Oracle with LLMs}:
In the remainder of this section, we give two examples that illustrate ways in which, even with a highly parsimonious pull request,
we are able to glean sufficient information to weed out false positives in both putative regression catching and  functionality catching JiTTests.

\subsection{Weeding out Fake Regression Catching}
Suppose we have a hardening JiTTest, $RCJ$, that we are considering as a potential Regression Catching JiTTest.

Since we consider it to be a regression catching test, then, by Definition~\ref{dfn:weak-regression-catching}, $RCJ$
passes on $\parent{TSIA}$ and fails on $TSIA$.
Now suppose that $RCJ$ asserts that the result of a query in the executed code returns a {\tt NULL} field for the location column. 
Such a test is seeking to {\em enforce} precisely the behaviour that we might reasonably infer $TSIA$ is seeking to extend/change.
If we present this test to the pull request author as a `catch'  we are likely to significantly increase the friction in their development experience. 
They will be understandably frustrated that their pull request, title `said it all', and our generated test paid little attention to this.

Even with this extremely parsimonious pull request, there is enough information in the title alone to automatically determine that $RCJ$ {\em should} be discarded, and its signal {\em not} passed to the engineer (unless it is to confirm that they have changed the functionality in the way they intended).
Therefore, an automated technique {\em would} be able to identify that the $RCJ$ test failure is likely to be a false positive, 
even with only those six words of the $TSIA$ pull request title and the behaviour of  $RCJ$ as guidance.

\subsection{Weeding out Fake Functionality Catching}
Suppose the code that implements the parsimonious pull request $TSIA$ includes logic that calculates the fuel consumed per distance travelled, 
as a  report on fuel efficiency.
Suppose that this logic is computed in terms of the location, and that the code in the pull request, relevant to this computation, is as follows.

{
\scriptsize
\begin{verbatim}
# Fuel efficiency computation assumes the destination 
# location cannot be the same as the current location
distance = distance(current_location, destination)
fuel_efficiency = fuel_consumed/distance
\end{verbatim}
}
Suppose further that the pull request also includes the implementation of the new function {\tt distance}.

Now suppose that we automatically generate a test $FCJ$ that, on the face of it, appears to be a functionality catching JiTTest.
By Definition ~\ref{dfn:weak-functionality-catching},  this means that  $FCJ$  fails to build (or builds yet fails) on the parent, and also fails on the pull request.
The new test, $FCJ$, calls the new logic to determine the distance and, therefore, does not build on the parent.
So far, so good: the test appears to be a functionality catching JiTTest.

Now comes the challenge: in order to be sure that the failure on $TSIA$ is a true positive, 
we need to scavenge as much oracle information as possible.
Suppose $FCJ$ fails with a divide by zero exception in the computation of {\tt  fuel\_efficiency}, and that it does this because the
current and destination locations turn out to be identical in the generated test case.
In this situation, the comment that immediately proceeds the computation of distance in the pull request is a highly relevant clue.
An automated test technique, especially one with the power of Large Language Model, 
{\em ought} to be able to infer, from this clue,  that $FCJ$ is failing precisely because it does  not respect the assumption captured by this comment.

Notice that this does not mean that we should not report {\em any} signal to the engineer.
Rather,  it would be better to construct a comment on the
pull request that observes that the test fails because the assumption is not met.

\noindent
{\bf Test signal refinement $\rightarrow$ Test Improvement}:
Furthermore, the tool could transform the test into a {\em passing} 
test that documents the assumption by asserting that the distance computation 
will raise a divide-by-zero exception when the current and destination location are identical.
This might encourage the engineer to make an improvement that avoids an exception.

\noindent
{\bf Test Improvement $\rightarrow$ Automated Repair}:
The automated system could also {\em suggest}  such an improvement:
test generation  thereby leading to improvement/repair suggestion.

\noindent
{\bf LLM + non-executable text = Oracle}:
As these two simple examples illustrate, the Catching JiTTest challenge is demanding, 
but not insurmountably so. 
In particular, we believe that LLMs' ability to reason with both code and natural 
language (in the same inference) provides a perfect bridge between the code and any available oracle information in non-executable text such as comments and other documentation, parsimonious though it may be. 
We see grounds for optimism in the use of `{\em oracle scavenging}'; the hunt for non-executable code hints to expected behaviour.
There has been much recent work on LLMs for testing~\cite{mhetal:LLM-survey,wang:LLM-testing-survey}, including 
recent work on the oracle inference problem~\cite{molina:test,fan:oracle,hossain:togll,konstantinou:llms},
but 
their potential to combine with test generation to tackle the  Catching JiTTest Challenge remains currently open and untackled,
yet highly impactful.

\subsection{Oracle Scavenging for Non-executable Code}
Non-executable text is primarily  authored by humans, 
and it will increasingly become the way in which humans drizzle oracle information into software.
This human author is arguably the best author for  such documentation, 
because human authors can provide these crucial fragments of oracle information.
As discussed in Section~\ref{sec:oracle}, Software Engineering is an inherently {\em social} process.
The oracle 
is one of the key points at which the social nature of the engineering discipline impinges.
Different humans may even disagree about what the expected behaviour of the software system should be.

The oracle may change over time, even with the same engineer changing their view of expected behaviour over time.
Regulatory and other changes may also impact the expected behaviour of software.
Therefore, without {\em any} further change in the code, 
the oracle may change.
It will become increasingly important, 
as language models play an increasingly wide role in code generation, 
to efficiently and accurately capture oracle information.

As the charter for the Source Code Analysis and Manipulation workshop points out:

\begin{quote}
``Source code contains the only precise description of the behaviour of the system''~\cite{scam-charter}.
\end{quote}

This makes source code analysis and manipulation extremely important, and it always will~\cite{mh:scam10-keynote},
but it also highlights what source code {\em cannot} do, 
as well as what it can do.
It is the only precise description of behaviour the system, 
but it is only that: 
a description of the behaviour of the system, 
not a description of its intended behaviour.
For the purpose of oracle extraction, 
we have to treat executable text (the executable source code) as potentially suspicious, 
and focus on extraction of intended semantics from non-executable text, 
so that we can compare intended and actual semantics.
This non-executable text may also be suspicious, of course, but at least it provides an alternative view of expected behaviour.
In the near future, there will be equal/greater importance for non-executable system descriptions, so
programming languages will need to evolve to support this. 

Perhaps we may even see a resurgence of interest in literate programming \cite{knuth:literate}.
In literate programming, the aim of non-executable text was to describe {\em how} the executable code performs its task.
Programming languages will need to evolve further to also include at least some informal specification of {\em what} those tasks are; what is the intended behaviour of the system.
We find ourselves back in previously well-trodden territory: agonizing over the absence of specifications and the relationship between specification and testing.
Formal specifications can help with test generation~\cite{rmhetal:fortest-survey}, but the kind of oracle information/specification required need not be formal.
Furthermore, the uncertainty about the true intended behaviour suggests that it would not even be desirable to have to rely on a purely formal specification.

\subsection{JiTTests Can Also Find Latent Bugs}
\label{sec:latent}
Many organisations have large legacy code bases and systems that contain 
latent bugs that have resided in the code base, sometimes for years.
By definition, such latent bugs are not mission-critical, since they have survived some time without correction. 
Nevertheless, although not mission-critical, they may, collectively, pose significant problems for users and can also unnecessarily drain resources and computational capacity.

Furthermore, some of these latent bugs may currently benefit from failed error propagation ~\cite{kaetal:analysis}.
In this failed error propagation situation, the bug is executed and `infects' the local state of computation, 
but this infection never manifests itself as a failure, 
because the bug is suppressed along all paths between the infected state and some point at which it becomes observable.
Such failed error propagation scenarios are vulnerable to future pull requests, that may inadvertently unblock a path from the infection point to an observation point, whereupon the latent bug newly leads to a failure; possibly a mission-critical failure.
In this regard, such latent bugs which currently benefit from failed error propagation are the software equivalent of unexploded bombs in the code base, they simply await detonation.

A Catching JiTTest is `just-in-time' for the pull request that it tests.
It might therefore appear that, by definition,  it cannot have a role to play in catching latent bugs,
but this is not the case.
We can take an arbitrary section of code of interest, delete it, and then re-insert it as a new pull request.
For example, suppose we pick a particular method, $m$. 
We delete $m$ and subsequently re-insert it as a new pull request. 
Suppose a Functionality Catching JiTTest, $t$, is generated for the pull request.
The test $t$ will thereby reveal a latent bug in $m$, that has already landed into production.

\noindent
{\bf Oracle Scavenging}:
The challenge of finding latent bugs with JiTTests lies primarily in locating additional sources of oracle information that help to determine 
what a re-inserted method (or code fragment)  $m$ is {\em supposed} to do.
When we seek to find latent bugs we have the most parsimonious pull request of all, 
from which we can scavenge no new non-executable text.
Fortunately, even for this  most challenging form of oracle scavenging, 
we can still automatically scour the continuous integration history for relevant past pull requests, 
extracting previous non-executable text from them. 
We can also seek out other documentation in the form of comments and other sources of non-executable text.
Indeed, this new world of LLM-based code generation (and test generation) 
dramatically increases the importance of such non-executable text: 
it is no longer just written for human consumption, it must  also be written for machine consumption.

No matter what way we seek to incorporate oracle information into the test generation approach, we will be able to use
JiTTest generation as an approach to uncover latent faults in existing software systems.
We can simply apply the delete-and-re-insert process arbitrarily many times to walk over a code base, using
JiTTest generation to uncover latent bugs.


\section{Conclusions and Future work}
This paper explores the complex interplay between test generation, test deployment and  test oracles.
Previous work has tended to focus solely on test generation, and the regression oracle.
We show that modes of test deployment have a profound impact on the way we seek to automatically generate tests and report their signal to engineers.
Our analysis also reveals the impact of taking non-regression oracle fragments into account.

Our primary contribution is to formally define and crystallise the research challenges that emerge from
this three-way interplay.
We hope that the paper clarifies the impact that this research agenda can have on industrial software testing. 
Our aim is to stimulate the research community to take up and investigate the challenges raised herein,
underpinned by precise formal definitions of the challenges involved, 
and motivated by industrial experience that highlights their potential impact.

\section*{Acknowledgements} 
We  would like to thank the leadership of Meta's Product Compliance and Privacy, Fundamental Artificial Intelligence Research (FAIR), Developer Infrastructure (DevInfra), and Instagram Product Foundation teams for supporting our work over the past decade, and the many Meta software engineers and testers whose experience, expertise, and engagement have helped  shape the ideas presented here. 
We would also like to thank the many academics and other researchers with whom we have had the enormous pleasure to interact over 3+ decades of research work on verification, testing and AI.

\newpage
\balance
%
\bibliographystyle{ACM-Reference-Format}
\bibliography{slice}


\begin{thebibliography}{86}


\ifx \showCODEN    \undefined \def \showCODEN     #1{\unskip}     \fi
\ifx \showDOI      \undefined \def \showDOI       #1{#1}\fi
\ifx \showISBNx    \undefined \def \showISBNx     #1{\unskip}     \fi
\ifx \showISBNxiii \undefined \def \showISBNxiii  #1{\unskip}     \fi
\ifx \showISSN     \undefined \def \showISSN      #1{\unskip}     \fi
\ifx \showLCCN     \undefined \def \showLCCN      #1{\unskip}     \fi
\ifx \shownote     \undefined \def \shownote      #1{#1}          \fi
\ifx \showarticletitle \undefined \def \showarticletitle #1{#1}   \fi
\ifx \showURL      \undefined \def \showURL       {\relax}        \fi
\providecommand\bibfield[2]{#2}
\providecommand\bibinfo[2]{#2}
\providecommand\natexlab[1]{#1}
\providecommand\showeprint[2][]{arXiv:#2}

\bibitem[Ahlgren et~al\mbox{.}(2020)]%
        {jaetal:gi20-keynote}
\bibfield{author}{\bibinfo{person}{John Ahlgren}, \bibinfo{person}{Maria~Eugenia Berezin}, \bibinfo{person}{Kinga Bojarczuk}, \bibinfo{person}{Elena Dulskyte}, \bibinfo{person}{Inna Dvortsova}, \bibinfo{person}{Johann George}, \bibinfo{person}{Natalija Gucevska}, \bibinfo{person}{Mark Harman}, \bibinfo{person}{Ralf Laemmel}, \bibinfo{person}{Erik Meijer}, \bibinfo{person}{Silvia Sapora}, {and} \bibinfo{person}{Justin Spahr-Summers}.} \bibinfo{year}{2020}\natexlab{}.
\newblock \showarticletitle{{WES}: Agent-based User Interaction Simulation on Real Infrastructure (keynote paper)}. In \bibinfo{booktitle}{\emph{$8^{th}$ Genetic improvement workshop ({GI at ICSE} 2020)}}, \bibfield{editor}{\bibinfo{person}{Shin Yoo}, \bibinfo{person}{Justyna Petke}, \bibinfo{person}{Westley Weimer}, {and} \bibinfo{person}{Bobby~R. Bruce}} (Eds.). \bibinfo{publisher}{ACM}, \bibinfo{pages}{276--284}.
\newblock


\bibitem[Ahlgren et~al\mbox{.}(2021a)]%
        {jaetal:mia}
\bibfield{author}{\bibinfo{person}{John Ahlgren}, \bibinfo{person}{Maria~Eugenia Berezin}, \bibinfo{person}{Kinga Bojarczuk}, \bibinfo{person}{Elena Dulskyte}, \bibinfo{person}{Inna Dvortsova}, \bibinfo{person}{Johann George}, \bibinfo{person}{Natalija Gucevska}, \bibinfo{person}{Mark Harman}, \bibinfo{person}{Maria Lomeli}, \bibinfo{person}{Erik Meijer}, \bibinfo{person}{Silvia Sapora}, {and} \bibinfo{person}{Justin Spahr-Summers}.} \bibinfo{year}{2021}\natexlab{a}.
\newblock \showarticletitle{Testing Web Enabled Simulation at Scale Using Metamorphic Testing}. In \bibinfo{booktitle}{\emph{International Conference on Software Engineering ({ICSE}) Software Engineering in Practice ({SEIP}) track}}. \bibinfo{address}{Virtual}.
\newblock


\bibitem[Ahlgren et~al\mbox{.}(2021b)]%
        {jaetal:ease21-keynote}
\bibfield{author}{\bibinfo{person}{John Ahlgren}, \bibinfo{person}{Kinga Bojarczuk}, \bibinfo{person}{Sophia Drossopoulou}, \bibinfo{person}{Inna Dvortsova}, \bibinfo{person}{Johann George}, \bibinfo{person}{Natalija Gucevska}, \bibinfo{person}{Mark Harman}, \bibinfo{person}{Maria Lomeli}, \bibinfo{person}{Simon Lucas}, \bibinfo{person}{Erik Meijer}, \bibinfo{person}{Steve Omohundro}, \bibinfo{person}{Rubmary Rojas}, \bibinfo{person}{Silvia Sapora}, \bibinfo{person}{Jie~M. Zhang}, {and} \bibinfo{person}{Norm Zhou}.} \bibinfo{year}{2021}\natexlab{b}.
\newblock \showarticletitle{Facebook's Cyber--Cyber and Cyber--Physical Digital Twins (keynote paper)}. In \bibinfo{booktitle}{\emph{25th International Conference on Evaluation and Assessment in Software Engineering ({EASE 2021})}}. \bibinfo{address}{Virtual}.
\newblock


\bibitem[Alshahwan et~al\mbox{.}(2024a)]%
        {kinga:enhancing}
\bibfield{author}{\bibinfo{person}{Nadia Alshahwan}, \bibinfo{person}{Arianna Blasi}, \bibinfo{person}{Kinga Bojarczuk}, \bibinfo{person}{Andrea Ciancone}, \bibinfo{person}{Natalija Gucevska}, \bibinfo{person}{Mark Harman}, \bibinfo{person}{Michal Krolikowski}, \bibinfo{person}{Rubmary Rojas}, \bibinfo{person}{Dragos Martac}, \bibinfo{person}{Simon Schellaert}, {et~al\mbox{.}}} \bibinfo{year}{2024}\natexlab{a}.
\newblock \showarticletitle{Enhancing Testing at {Meta} with Rich-State Simulated Populations}. In \bibinfo{booktitle}{\emph{Proceedings of the 46th International Conference on Software Engineering: Software Engineering in Practice}}. \bibinfo{pages}{1--12}.
\newblock


\bibitem[Alshahwan et~al\mbox{.}(2024b)]%
        {mhetal:TestGen-LLM}
\bibfield{author}{\bibinfo{person}{Nadia Alshahwan}, \bibinfo{person}{Jubin Chheda}, \bibinfo{person}{Anastasia Finegenova}, \bibinfo{person}{Mark Harman}, \bibinfo{person}{Alexandru Marginean}, \bibinfo{person}{Shubho Sengupta}, {and} \bibinfo{person}{Eddy Wang}.} \bibinfo{year}{2024}\natexlab{b}.
\newblock \showarticletitle{Automated unit test improvement using {Large Language Models} at {Meta}}. In \bibinfo{booktitle}{\emph{{ACM} International Conference on the Foundations of Software Engineering ({FSE 2024})}} (Porto de Galinhas, Brazil, Brazil).
\newblock


\bibitem[Alshahwan et~al\mbox{.}(2018)]%
        {mhetal:ssbse18-keynote}
\bibfield{author}{\bibinfo{person}{Nadia Alshahwan}, \bibinfo{person}{Xinbo Gao}, \bibinfo{person}{Mark Harman}, \bibinfo{person}{Yue Jia}, \bibinfo{person}{Ke Mao}, \bibinfo{person}{Alexander Mols}, \bibinfo{person}{Taijin Tei}, {and} \bibinfo{person}{Ilya Zorin}.} \bibinfo{year}{2018}\natexlab{}.
\newblock \showarticletitle{Deploying Search Based Software Engineering with {S}apienz at {F}acebook (keynote paper)}. In \bibinfo{booktitle}{\emph{$10^{th}$ International Symposium on Search Based Software Engineering ({SSBSE 2018})}}. \bibinfo{address}{{M}ontpellier, {F}rance}, \bibinfo{pages}{3--45}.
\newblock
\newblock
\shownote{Springer {LNCS} 11036}.


\bibitem[Alshahwan et~al\mbox{.}(2023)]%
        {alshahwan:software}
\bibfield{author}{\bibinfo{person}{Nadia Alshahwan}, \bibinfo{person}{Mark Harman}, {and} \bibinfo{person}{Alexandru Marginean}.} \bibinfo{year}{2023}\natexlab{}.
\newblock \showarticletitle{Software Testing Research Challenges: An Industrial Perspective (keynote paper)}. In \bibinfo{booktitle}{\emph{2023 {IEEE} Conference on Software Testing, Verification and Validation ({ICST 2023})}}. IEEE, \bibinfo{pages}{1--10}.
\newblock


\bibitem[Alshahwan et~al\mbox{.}(2024d)]%
        {mhetal:intense24-keynote}
\bibfield{author}{\bibinfo{person}{Nadia Alshahwan}, \bibinfo{person}{Mark Harman}, \bibinfo{person}{Alexandru Marginean}, \bibinfo{person}{Shubho Sengupta}, {and} \bibinfo{person}{Eddy Wang}.} \bibinfo{year}{2024}\natexlab{d}.
\newblock \showarticletitle{Assured LLM-Based Software Engineering (keynote paper)}. In \bibinfo{booktitle}{\emph{$2^{nd.}$ {ICSE} workshop on Interoperability and Robustness of Neural Software Engineering ({InteNSE})}} (Lisbon, Portugal).
\newblock


\bibitem[Alshahwan et~al\mbox{.}(2024c)]%
        {mhetal:TestGen-obs}
\bibfield{author}{\bibinfo{person}{Nadia Alshahwan}, \bibinfo{person}{Mark Harman}, \bibinfo{person}{Alexandru Marginean}, {and} \bibinfo{person}{Eddy Wang}.} \bibinfo{year}{2024}\natexlab{c}.
\newblock \showarticletitle{Observation-based unit test generation at Meta}. In \bibinfo{booktitle}{\emph{Foundations of Software Engineering ({FSE} 2024)}}.
\newblock


\bibitem[Analysis and ({SCAM})(2001)]%
        {scam-charter}
\bibfield{author}{\bibinfo{person}{Source~Code Analysis} {and} \bibinfo{person}{Manipulation~Workshop ({SCAM})}.} \bibinfo{year}{2001}\natexlab{}.
\newblock \bibinfo{title}{{SCAM Charter}}.
\newblock
\newblock
\urldef\tempurl%
\url{https://www.ieee-scam.org/}
\showURL{%
\tempurl}
\newblock
\shownote{The cited quotation has remained part of the workshop's credo since its foundation in 2001}.


\bibitem[Anand et~al\mbox{.}(2013)]%
        {anandetal:orchestrated}
\bibfield{author}{\bibinfo{person}{Saswat Anand}, \bibinfo{person}{Antonia Bertolino}, \bibinfo{person}{Edmund Burke}, \bibinfo{person}{Tsong~Yueh Chen}, \bibinfo{person}{John Clark}, \bibinfo{person}{Myra~B. Cohen}, \bibinfo{person}{Wolfgang Grieskamp}, \bibinfo{person}{Mark Harman}, \bibinfo{person}{Mary~Jean Harrold}, \bibinfo{person}{Jenny Li}, \bibinfo{person}{Phil McMinn}, {and} \bibinfo{person}{Hong Zhu}.} \bibinfo{year}{2013}\natexlab{}.
\newblock \showarticletitle{An orchestrated survey of methodologies for automated software test case generation}.
\newblock \bibinfo{journal}{\emph{Journal of Systems and Software}} \bibinfo{volume}{86}, \bibinfo{number}{8} (\bibinfo{date}{August} \bibinfo{year}{2013}), \bibinfo{pages}{1978--2001}.
\newblock


\bibitem[Androutsopoulos et~al\mbox{.}(2014)]%
        {kaetal:analysis}
\bibfield{author}{\bibinfo{person}{Kelly Androutsopoulos}, \bibinfo{person}{David Clark}, \bibinfo{person}{Haitao Dan}, \bibinfo{person}{Mark Harman}, {and} \bibinfo{person}{Robert Hierons}.} \bibinfo{year}{2014}\natexlab{}.
\newblock \showarticletitle{An Analysis of the Relationship between Conditional Entropy and Failed Error Propagation in Software Testing}. In \bibinfo{booktitle}{\emph{$36^{th}$ International Conference on Software Engineering ({ICSE 2014})}}. \bibinfo{address}{Hyderabad, India}, \bibinfo{pages}{573--583}.
\newblock


\bibitem[Barr et~al\mbox{.}(2015)]%
        {ebetal:oracle}
\bibfield{author}{\bibinfo{person}{Earl~T. Barr}, \bibinfo{person}{Mark Harman}, \bibinfo{person}{Phil McMinn}, \bibinfo{person}{Muzammil Shahbaz}, {and} \bibinfo{person}{Shin Yoo}.} \bibinfo{year}{2015}\natexlab{}.
\newblock \showarticletitle{The Oracle Problem in Software Testing: {A} Survey}.
\newblock \bibinfo{journal}{\emph{{IEEE} Transactions on Software Engineering}} \bibinfo{volume}{41}, \bibinfo{number}{5} (\bibinfo{date}{May} \bibinfo{year}{2015}), \bibinfo{pages}{507--525}.
\newblock


\bibitem[Brandt and Zaidman(2022)]%
        {brandt2022developer}
\bibfield{author}{\bibinfo{person}{Carolin Brandt} {and} \bibinfo{person}{Andy Zaidman}.} \bibinfo{year}{2022}\natexlab{}.
\newblock \showarticletitle{Developer-centric test amplification: The interplay between automatic generation human exploration}.
\newblock \bibinfo{journal}{\emph{Empirical Software Engineering}} \bibinfo{volume}{27}, \bibinfo{number}{4} (\bibinfo{year}{2022}), \bibinfo{pages}{96}.
\newblock


\bibitem[Cadar et~al\mbox{.}(2008)]%
        {cadar:klee}
\bibfield{author}{\bibinfo{person}{Cristian Cadar}, \bibinfo{person}{Daniel Dunbar}, \bibinfo{person}{Dawson~R Engler}, {et~al\mbox{.}}} \bibinfo{year}{2008}\natexlab{}.
\newblock \showarticletitle{Klee: unassisted and automatic generation of high-coverage tests for complex systems programs.}. In \bibinfo{booktitle}{\emph{OSDI}}, Vol.~\bibinfo{volume}{8}. \bibinfo{pages}{209--224}.
\newblock


\bibitem[Cadar et~al\mbox{.}(2011)]%
        {cadar:icse11}
\bibfield{author}{\bibinfo{person}{Cristian Cadar}, \bibinfo{person}{Patrice Godefroid}, \bibinfo{person}{Sarfraz Khurshid}, \bibinfo{person}{Corina~S. P\u{a}s\u{a}reanu}, \bibinfo{person}{Koushik Sen}, \bibinfo{person}{Nikolai Tillmann}, {and} \bibinfo{person}{Willem Visser}.} \bibinfo{year}{2011}\natexlab{}.
\newblock \showarticletitle{Symbolic execution for software testing in practice: preliminary assessment}. In \bibinfo{booktitle}{\emph{$33^{rd}$ International Conference on Software Engineering ({ICSE'11})}} (Waikiki, Honolulu, HI, USA). \bibinfo{publisher}{ACM}, \bibinfo{address}{New York, NY, USA}, \bibinfo{pages}{1066--1071}.
\newblock
\showISBNx{978-1-4503-0445-0}


\bibitem[Cadar and Sen(2013)]%
        {cadar:three-decades}
\bibfield{author}{\bibinfo{person}{Cristian Cadar} {and} \bibinfo{person}{Koushik Sen}.} \bibinfo{year}{2013}\natexlab{}.
\newblock \showarticletitle{Symbolic Execution for Software Testing: Three Decades Later}.
\newblock \bibinfo{journal}{\emph{Commun. ACM}} \bibinfo{volume}{56}, \bibinfo{number}{2} (\bibinfo{date}{Feb.} \bibinfo{year}{2013}), \bibinfo{pages}{82--90}.
\newblock
\showISSN{0001-0782}


\bibitem[Calcagno et~al\mbox{.}(2015)]%
        {movefast}
\bibfield{author}{\bibinfo{person}{C. Calcagno}, \bibinfo{person}{D. Distefano}, \bibinfo{person}{J. Dubreil}, \bibinfo{person}{D. Gabi}, \bibinfo{person}{P. Hooimeijer}, \bibinfo{person}{M. Luca}, \bibinfo{person}{P.~W. O'Hearn}, \bibinfo{person}{I. Papakonstantinou}, \bibinfo{person}{J. Purbrick}, {and} \bibinfo{person}{D. Rodriguez}.} \bibinfo{year}{2015}\natexlab{}.
\newblock \showarticletitle{Moving Fast with Software Verification}. In \bibinfo{booktitle}{\emph{{NASA} Formal Methods - 7th International Symposium}}. \bibinfo{pages}{3--11}.
\newblock


\bibitem[Chekam et~al\mbox{.}(2017)]%
        {mike:icse17}
\bibfield{author}{\bibinfo{person}{Thierry~Titcheu Chekam}, \bibinfo{person}{Mike Papadakis}, \bibinfo{person}{Yves~Le Traon}, {and} \bibinfo{person}{Mark Harman}.} \bibinfo{year}{2017}\natexlab{}.
\newblock \showarticletitle{An empirical study on mutation, statement and branch coverage fault revelation that avoids the unreliable clean program assumption}. In \bibinfo{booktitle}{\emph{Proceedings of the 39th International Conference on Software Engineering, {ICSE} 2017, Buenos Aires, Argentina, May 20-28, 2017}}. \bibinfo{pages}{597--608}.
\newblock


\bibitem[Cordy et~al\mbox{.}(2022)]%
        {mcetal:flakime}
\bibfield{author}{\bibinfo{person}{Maxime Cordy}, \bibinfo{person}{Renaud Rwemalika}, \bibinfo{person}{Adriano Franci}, \bibinfo{person}{Mike Papadakis}, {and} \bibinfo{person}{Mark Harman}.} \bibinfo{year}{2022}\natexlab{}.
\newblock \showarticletitle{FlakiMe: Laboratory-Controlled Test Flakiness Impact Assessment}. In \bibinfo{booktitle}{\emph{44th {IEEE/ACM} 44th International Conference on Software Engineering, {ICSE} 2022, Pittsburgh, PA, USA, May 25-27, 2022}}. \bibinfo{publisher}{{ACM}}, \bibinfo{pages}{982--994}.
\newblock


\bibitem[{De Millo} et~al\mbox{.}(1979)]%
        {demillo:social}
\bibfield{author}{\bibinfo{person}{Richard~A. {De Millo}}, \bibinfo{person}{Richard~J. Lipton}, {and} \bibinfo{person}{Alan~J. Perlis}.} \bibinfo{year}{1979}\natexlab{}.
\newblock \showarticletitle{Social Processes and Proofs of Theorems and Programs}.
\newblock \bibinfo{journal}{\emph{Commun. ACM}} \bibinfo{volume}{22}, \bibinfo{number}{5} (\bibinfo{date}{May} \bibinfo{year}{1979}), \bibinfo{pages}{271--280}.
\newblock
\newblock
\shownote{An earlier version appeared in \bgroup\em {ACM} {S}ymposium on {P}rinciples of {P}rogramming {L}anguages ({POPL}) , Los Angeles, California \egroup, 1977 pp. 206--214}.


\bibitem[Dijkstra(1978)]%
        {dijkstra:pamphlet}
\bibfield{author}{\bibinfo{person}{Edsger~W. Dijkstra}.} \bibinfo{year}{1978}\natexlab{}.
\newblock \showarticletitle{On a Political Pamphlet from the Middle Ages ({A} response to the paper `Social Processes and Proofs of Theorems and Programs' by {DeMillo}, {Lipton}, and {Perlis})}.
\newblock \bibinfo{journal}{\emph{{ACM} {SIGSOFT}, {S}oftware Engineering Notes}} \bibinfo{volume}{3}, \bibinfo{number}{2} (\bibinfo{year}{1978}), \bibinfo{pages}{14--17}.
\newblock


\bibitem[Distefano et~al\mbox{.}(2019)]%
        {distefano:scaling}
\bibfield{author}{\bibinfo{person}{Dino Distefano}, \bibinfo{person}{Manuel F{\"a}hndrich}, \bibinfo{person}{Francesco Logozzo}, {and} \bibinfo{person}{Peter~W O'Hearn}.} \bibinfo{year}{2019}\natexlab{}.
\newblock \showarticletitle{Scaling static analyses at Facebook}.
\newblock \bibinfo{journal}{\emph{Commun. ACM}} \bibinfo{volume}{62}, \bibinfo{number}{8} (\bibinfo{year}{2019}), \bibinfo{pages}{62--70}.
\newblock


\bibitem[Dodhiawala et~al\mbox{.}(1989)]%
        {dodhiawala:real}
\bibfield{author}{\bibinfo{person}{Rajendra~T Dodhiawala}, \bibinfo{person}{NS Sridharan}, \bibinfo{person}{Peter Raulefs}, {and} \bibinfo{person}{Cynthia Pickering}.} \bibinfo{year}{1989}\natexlab{}.
\newblock \showarticletitle{Real-Time {AI} Systems: {A} Definition and An Architecture.}. In \bibinfo{booktitle}{\emph{IJCAI}}. Citeseer, \bibinfo{pages}{256--264}.
\newblock


\bibitem[Elbaum et~al\mbox{.}(2009)]%
        {elbaum:carving}
\bibfield{author}{\bibinfo{person}{Sebastian~G. Elbaum}, \bibinfo{person}{Hui~Nee Chin}, \bibinfo{person}{Matthew~B. Dwyer}, {and} \bibinfo{person}{Matthew Jorde}.} \bibinfo{year}{2009}\natexlab{}.
\newblock \showarticletitle{Carving and Replaying Differential Unit Test Cases from System Test Cases}.
\newblock \bibinfo{journal}{\emph{IEEE Transactions on Software Engineering}} \bibinfo{volume}{35}, \bibinfo{number}{1} (\bibinfo{year}{2009}), \bibinfo{pages}{29--45}.
\newblock


\bibitem[Fan et~al\mbox{.}(2023)]%
        {mhetal:LLM-survey}
\bibfield{author}{\bibinfo{person}{Angela Fan}, \bibinfo{person}{Beliz Gokkaya}, \bibinfo{person}{Mitya Lyubarskiy}, \bibinfo{person}{Mark Harman}, \bibinfo{person}{Shubho Sengupta}, \bibinfo{person}{Shin Yoo}, {and} \bibinfo{person}{Jie Zhang}.} \bibinfo{year}{2023}\natexlab{}.
\newblock \showarticletitle{{L}arge {L}anguage {M}odels for {S}oftware {E}ngineering: {S}urvey and Open Problems}. In \bibinfo{booktitle}{\emph{{ICSE} {F}uture of {S}oftware {E}ngineering ({FoSE} 2023)}}.
\newblock


\bibitem[Fan et~al\mbox{.}(2024)]%
        {fan:oracle}
\bibfield{author}{\bibinfo{person}{Zhiyu Fan}, \bibinfo{person}{Haifeng Ruan}, \bibinfo{person}{Sergey Mechtaev}, {and} \bibinfo{person}{Abhik Roychoudhury}.} \bibinfo{year}{2024}\natexlab{}.
\newblock \showarticletitle{Oracle-guided Program Selection from {L}arge {L}anguage {M}odels}. In \bibinfo{booktitle}{\emph{Proceedings of the 33rd {ACM SIGSOFT} International Symposium on Software Testing and Analysis}}. \bibinfo{pages}{628--640}.
\newblock


\bibitem[Fernandez et~al\mbox{.}(1996)]%
        {fernandez1996using}
\bibfield{author}{\bibinfo{person}{Jean~Claude Fernandez}, \bibinfo{person}{Claude Jard}, \bibinfo{person}{Thierry J{\'e}ron}, {and} \bibinfo{person}{C{\'e}sar Viho}.} \bibinfo{year}{1996}\natexlab{}.
\newblock \showarticletitle{Using on-the-fly verification techniques for the generation of test suites}. In \bibinfo{booktitle}{\emph{Computer Aided Verification: 8th International Conference, CAV'96 New Brunswick, NJ, USA, July 31--August 3, 1996 Proceedings 8}}. Springer, \bibinfo{pages}{348--359}.
\newblock


\bibitem[Foster et~al\mbox{.}(2025)]%
        {foster:mutation}
\bibfield{author}{\bibinfo{person}{Christopher Foster}, \bibinfo{person}{Abhishek Gulati}, \bibinfo{person}{Mark Harman}, \bibinfo{person}{Inna Harper}, \bibinfo{person}{Ke Mao}, \bibinfo{person}{Jillian Ritchey}, \bibinfo{person}{Herv{\'e} Robert}, {and} \bibinfo{person}{Shubho Sengupta}.} \bibinfo{year}{2025}\natexlab{}.
\newblock \showarticletitle{Mutation-Guided LLM-based Test Generation at Meta}. In \bibinfo{booktitle}{\emph{2025 {ACM} Conference on Foundations of Software Engineering ({FSE 2025})}}. {ACM}.
\newblock
\newblock
\shownote{Also available as arXiv preprint arXiv:2501.12862}.


\bibitem[Frankl et~al\mbox{.}(1997)]%
        {frankl_etal97}
\bibfield{author}{\bibinfo{person}{Phyllis~G. Frankl}, \bibinfo{person}{Stewart~N. Weiss}, {and} \bibinfo{person}{Cang Hu}.} \bibinfo{year}{1997}\natexlab{}.
\newblock \showarticletitle{All-Uses vs Mutation Testing: An Experimental Comparison of Effectiveness}.
\newblock \bibinfo{journal}{\emph{{J}ournal of {S}ystems {S}oftware}}  \bibinfo{volume}{38} (\bibinfo{year}{1997}), \bibinfo{pages}{235--253}.
\newblock


\bibitem[Fraser and Arcuri(2011)]%
        {fraser:evosuite}
\bibfield{author}{\bibinfo{person}{Gordon Fraser} {and} \bibinfo{person}{Andrea Arcuri}.} \bibinfo{year}{2011}\natexlab{}.
\newblock \showarticletitle{{EvoSuite}: automatic test suite generation for object-oriented software}. In \bibinfo{booktitle}{\emph{$8^{th}$ European Software Engineering Conference and the {ACM SIGSOFT} Symposium on the Foundations of Software Engineering ({ESEC/FSE '11})}}. \bibinfo{publisher}{ACM}, \bibinfo{pages}{416--419}.
\newblock
\showISBNx{978-1-4503-0443-6}


\bibitem[Godefroid et~al\mbox{.}(2005)]%
        {godefroid:dart}
\bibfield{author}{\bibinfo{person}{Patrice Godefroid}, \bibinfo{person}{Nils Klarlund}, {and} \bibinfo{person}{Koushik Sen}.} \bibinfo{year}{2005}\natexlab{}.
\newblock \showarticletitle{{DART}: directed automated random testing}. In \bibinfo{booktitle}{\emph{Programming Language Design and Implementation ({PLDI 2005})}}, \bibfield{editor}{\bibinfo{person}{Vivek Sarkar} {and} \bibinfo{person}{Mary~W. Hall}} (Eds.). \bibinfo{publisher}{ACM}, \bibinfo{pages}{213--223}.
\newblock
\showISBNx{1-59593-056-6}


\bibitem[Gopinath et~al\mbox{.}(2014)]%
        {gopinath2014code}
\bibfield{author}{\bibinfo{person}{Rahul Gopinath}, \bibinfo{person}{Carlos Jensen}, {and} \bibinfo{person}{Alex Groce}.} \bibinfo{year}{2014}\natexlab{}.
\newblock \showarticletitle{Code coverage for suite evaluation by developers}. In \bibinfo{booktitle}{\emph{Proceedings of the 36th international conference on software engineering}}. \bibinfo{pages}{72--82}.
\newblock


\bibitem[Harman(2010)]%
        {mh:scam10-keynote}
\bibfield{author}{\bibinfo{person}{Mark Harman}.} \bibinfo{year}{2010}\natexlab{}.
\newblock \showarticletitle{Why Source Code Analysis and Manipulation Will Always Be Important (Keynote Paper)}. In \bibinfo{booktitle}{\emph{$10^{th}$ {IEEE} International Working Conference on Source Code Analysis and Manipulation}}. \bibinfo{address}{Timisoara, Romania}.
\newblock


\bibitem[Harman et~al\mbox{.}(2011)]%
        {mhetal:shom}
\bibfield{author}{\bibinfo{person}{Mark Harman}, \bibinfo{person}{Yue Jia}, {and} \bibinfo{person}{William~B. Langdon}.} \bibinfo{year}{2011}\natexlab{}.
\newblock \showarticletitle{Strong Higher Order Mutation-Based Test Data Generation}. In \bibinfo{booktitle}{\emph{$8^{th}$ European Software Engineering Conference and the {ACM SIGSOFT} Symposium on the Foundations of Software Engineering ({ESEC/FSE '11})}} (Szeged, Hungary). \bibinfo{publisher}{ACM}, \bibinfo{address}{New York, NY, USA}, \bibinfo{pages}{212--222}.
\newblock
\showISBNx{978-1-4503-0443-6}


\bibitem[Harman et~al\mbox{.}(2015)]%
        {mh:icst15-keynote}
\bibfield{author}{\bibinfo{person}{Mark Harman}, \bibinfo{person}{Yue Jia}, {and} \bibinfo{person}{Yuanyuan Zhang}.} \bibinfo{year}{2015}\natexlab{}.
\newblock \showarticletitle{Achievements, open problems and challenges for search based software testing (keynote Paper)}. In \bibinfo{booktitle}{\emph{$8^{th}$ {IEEE} International Conference on Software Testing, Verification and Validation ({ICST 2015})}}. \bibinfo{address}{Graz, Austria}.
\newblock


\bibitem[Harman and Jones(2001)]%
        {mhbj:manifesto}
\bibfield{author}{\bibinfo{person}{Mark Harman} {and} \bibinfo{person}{Bryan~F. Jones}.} \bibinfo{year}{2001}\natexlab{}.
\newblock \showarticletitle{Search Based Software Engineering}.
\newblock \bibinfo{journal}{\emph{{I}nformation and {S}oftware {T}echnology}} \bibinfo{volume}{43}, \bibinfo{number}{14} (\bibinfo{date}{Dec.} \bibinfo{year}{2001}), \bibinfo{pages}{833--839}.
\newblock
\showISSN{0950-5849}


\bibitem[Harman et~al\mbox{.}(2012)]%
        {mhamyz:acm-surveys}
\bibfield{author}{\bibinfo{person}{Mark Harman}, \bibinfo{person}{Afshin Mansouri}, {and} \bibinfo{person}{Yuanyuan Zhang}.} \bibinfo{year}{2012}\natexlab{}.
\newblock \showarticletitle{Search Based Software Engineering: {T}rends, Techniques and Applications}.
\newblock \bibinfo{journal}{\emph{Comput. Surveys}} \bibinfo{volume}{45}, \bibinfo{number}{1} (\bibinfo{date}{November} \bibinfo{year}{2012}), \bibinfo{pages}{11:1--11:61}.
\newblock


\bibitem[Harman and {O'H}earn(2018)]%
        {mhpoh:scam18-keynote}
\bibfield{author}{\bibinfo{person}{Mark Harman} {and} \bibinfo{person}{Peter {O'H}earn}.} \bibinfo{year}{2018}\natexlab{}.
\newblock \showarticletitle{From Start-ups to Scale-ups: {O}pportunities and Open Problems for Static and Dynamic Program Analysis (keynote paper)}. In \bibinfo{booktitle}{\emph{$18^{th}$ {IEEE} International Working Conference on Source Code Analysis and Manipulation ({SCAM 2018})}}. \bibinfo{address}{{M}adrid, {S}pain}, \bibinfo{pages}{1--23}.
\newblock


\bibitem[Harman and Tratt(2007)]%
        {laurie:gecco07}
\bibfield{author}{\bibinfo{person}{Mark Harman} {and} \bibinfo{person}{Laurence Tratt}.} \bibinfo{year}{2007}\natexlab{}.
\newblock \showarticletitle{Pareto Optimal Search-based Refactoring at the Design Level}. In \bibinfo{booktitle}{\emph{$9^{th}$ annual conference on Genetic and evolutionary computation ({GECCO 2007})}}. \bibinfo{publisher}{ACM Press}, \bibinfo{address}{London, {UK}}, \bibinfo{pages}{1106 -- 1113}.
\newblock


\bibitem[Harman et~al\mbox{.}(2014)]%
        {xymhyj:equivalent}
\bibfield{author}{\bibinfo{person}{Mark Harman}, \bibinfo{person}{Xiangjuan Yao}, {and} \bibinfo{person}{Yue Jia}.} \bibinfo{year}{2014}\natexlab{}.
\newblock \showarticletitle{A Study of Equivalent and Stubborn Mutation Operators Using Human Analysis of Equivalence}. In \bibinfo{booktitle}{\emph{$36^{th}$ International Conference on Software Engineering ({ICSE 2014})}}. \bibinfo{address}{Hyderabad, India}, \bibinfo{pages}{919--930}.
\newblock


\bibitem[Hasan et~al\mbox{.}(2023)]%
        {hasan2023understanding}
\bibfield{author}{\bibinfo{person}{Kazi~Amit Hasan}, \bibinfo{person}{Marcos Macedo}, \bibinfo{person}{Yuan Tian}, \bibinfo{person}{Bram Adams}, {and} \bibinfo{person}{Steven Ding}.} \bibinfo{year}{2023}\natexlab{}.
\newblock \showarticletitle{Understanding the time to first response in GitHub pull requests}. In \bibinfo{booktitle}{\emph{2023 IEEE/ACM 20th International Conference on Mining Software Repositories (MSR)}}. IEEE, \bibinfo{pages}{1--11}.
\newblock


\bibitem[Hierons et~al\mbox{.}(2009)]%
        {rmhetal:fortest-survey}
\bibfield{author}{\bibinfo{person}{Rob Hierons}, \bibinfo{person}{Kirill Bogdanov}, \bibinfo{person}{Jonathan Bowen}, \bibinfo{person}{Rance Cleaveland}, \bibinfo{person}{John Derrick}, \bibinfo{person}{Jeremy Dick}, \bibinfo{person}{Marian Gheorghe}, \bibinfo{person}{Mark Harman}, \bibinfo{person}{Kalpesh Kapoor}, \bibinfo{person}{Paul Krause}, \bibinfo{person}{Gerald Luettgen}, \bibinfo{person}{Tony Simons}, \bibinfo{person}{Sergiy Vilkomir}, \bibinfo{person}{Martin Woodward}, {and} \bibinfo{person}{Hussein Zedan}.} \bibinfo{year}{2009}\natexlab{}.
\newblock \showarticletitle{Using Formal Methods to Support Testing}.
\newblock \bibinfo{journal}{\emph{Comput. Surveys}} \bibinfo{volume}{41}, \bibinfo{number}{2} (\bibinfo{date}{Feb.} \bibinfo{year}{2009}).
\newblock
\newblock
\shownote{Article 9}.


\bibitem[Hoare(1996a)]%
        {hoare:how}
\bibfield{author}{\bibinfo{person}{Charles Anthony~Richard Hoare}.} \bibinfo{year}{1996}\natexlab{a}.
\newblock \showarticletitle{How did software get so reliable without proof?}. In \bibinfo{booktitle}{\emph{{FME '96}: Industrial Benefit and Advances in Formal Methods: Third International Symposium of Formal Methods Europe}} \emph{(\bibinfo{series}{LNCS}, \bibinfo{number}{1051})}. \bibinfo{publisher}{Springer-Verlag}, \bibinfo{pages}{1--17}.
\newblock


\bibitem[Hoare(1996b)]%
        {hoare:icse96}
\bibfield{author}{\bibinfo{person}{Charles Anthony~Richard Hoare}.} \bibinfo{year}{1996}\natexlab{b}.
\newblock \showarticletitle{How did software get so reliable without proof?}. In \bibinfo{booktitle}{\emph{{IEEE} International Conference on Software Engineering ({ICSE'96})}}. \bibinfo{publisher}{{IEEE} {C}omputer {S}ociety {P}ress}, \bibinfo{address}{Los Alamitos, California, {USA}}.
\newblock
\newblock
\shownote{Keynote talk and extended abstract}.


\bibitem[Hoi et~al\mbox{.}(2021)]%
        {hoi:online}
\bibfield{author}{\bibinfo{person}{Steven~CH Hoi}, \bibinfo{person}{Doyen Sahoo}, \bibinfo{person}{Jing Lu}, {and} \bibinfo{person}{Peilin Zhao}.} \bibinfo{year}{2021}\natexlab{}.
\newblock \showarticletitle{Online learning: {A} comprehensive survey}.
\newblock \bibinfo{journal}{\emph{Neurocomputing}}  \bibinfo{volume}{459} (\bibinfo{year}{2021}), \bibinfo{pages}{249--289}.
\newblock


\bibitem[Hossain and Dwyer(2024)]%
        {hossain:togll}
\bibfield{author}{\bibinfo{person}{Soneya~Binta Hossain} {and} \bibinfo{person}{Matthew Dwyer}.} \bibinfo{year}{2024}\natexlab{}.
\newblock \showarticletitle{{TOGLL}: Correct and strong test oracle generation with {LLMs}}.
\newblock \bibinfo{journal}{\emph{arXiv preprint arXiv:2405.03786}} (\bibinfo{year}{2024}).
\newblock


\bibitem[{Ie} et~al\mbox{.}(2019)]%
        {Ie:RecSym}
\bibfield{author}{\bibinfo{person}{Eugene {Ie}}, \bibinfo{person}{Chih-wei {Hsu}}, \bibinfo{person}{Martin {Mladenov}}, \bibinfo{person}{Vihan {Jain}}, \bibinfo{person}{Sanmit {Narvekar}}, \bibinfo{person}{Jing {Wang}}, \bibinfo{person}{Rui {Wu}}, {and} \bibinfo{person}{Craig {Boutilier}}.} \bibinfo{year}{2019}\natexlab{}.
\newblock \showarticletitle{{RecSim: A Configurable Simulation Platform for Recommender Systems}}.
\newblock \bibinfo{journal}{\emph{arXiv e-prints}}, Article \bibinfo{articleno}{arXiv:1909.04847} (\bibinfo{date}{Sep} \bibinfo{year}{2019}).
\newblock
\showeprint[arxiv]{1909.04847}~[cs.LG]


\bibitem[Inozemtseva and Holmes(2014)]%
        {inozemtseva2014coverage}
\bibfield{author}{\bibinfo{person}{Laura Inozemtseva} {and} \bibinfo{person}{Reid Holmes}.} \bibinfo{year}{2014}\natexlab{}.
\newblock \showarticletitle{Coverage is not strongly correlated with test suite effectiveness}. In \bibinfo{booktitle}{\emph{Proceedings of the 36th international conference on software engineering}}. \bibinfo{pages}{435--445}.
\newblock


\bibitem[Jain et~al\mbox{.}(2024)]%
        {jain2024testgeneval}
\bibfield{author}{\bibinfo{person}{Kush Jain}, \bibinfo{person}{Gabriel Synnaeve}, {and} \bibinfo{person}{Baptiste Rozi{\`e}re}.} \bibinfo{year}{2024}\natexlab{}.
\newblock \showarticletitle{{TestGenEval}: {A} real world unit test generation and test completion benchmark}.
\newblock \bibinfo{journal}{\emph{arXiv preprint arXiv:2410.00752}} (\bibinfo{year}{2024}).
\newblock


\bibitem[Jia and Harman(2011)]%
        {yjmh:analysis}
\bibfield{author}{\bibinfo{person}{Yue Jia} {and} \bibinfo{person}{Mark Harman}.} \bibinfo{year}{2011}\natexlab{}.
\newblock \showarticletitle{An Analysis and Survey of the Development of Mutation Testing}.
\newblock \bibinfo{journal}{\emph{{IEEE} Transactions on Software Engineering}} \bibinfo{volume}{37}, \bibinfo{number}{5} (\bibinfo{date}{September--October} \bibinfo{year}{2011}), \bibinfo{pages}{649 -- 678}.
\newblock


\bibitem[Just et~al\mbox{.}(2014)]%
        {just:are-mutants}
\bibfield{author}{\bibinfo{person}{Ren\'{e} Just}, \bibinfo{person}{Darioush Jalali}, \bibinfo{person}{Laura Inozemtseva}, \bibinfo{person}{Michael~D. Ernst}, \bibinfo{person}{Reid Holmes}, {and} \bibinfo{person}{Gordon Fraser}.} \bibinfo{year}{2014}\natexlab{}.
\newblock \bibinfo{booktitle}{\emph{Are Mutants a Valid Substitute for Real Faults in Software Testing?}}
\newblock \bibinfo{type}{{T}echnical {R}eport} {UW-CSE-14-02-02}. \bibinfo{institution}{University of Washington}.
\newblock


\bibitem[Knuth(1984)]%
        {knuth:literate}
\bibfield{author}{\bibinfo{person}{Donald~E. Knuth}.} \bibinfo{year}{1984}\natexlab{}.
\newblock \showarticletitle{Literate Programming}.
\newblock \bibinfo{journal}{\emph{Comput. J.}} \bibinfo{volume}{27}, \bibinfo{number}{2} (\bibinfo{year}{1984}), \bibinfo{pages}{97--111}.
\newblock


\bibitem[Konstantinou et~al\mbox{.}(2024)]%
        {konstantinou:llms}
\bibfield{author}{\bibinfo{person}{Michael Konstantinou}, \bibinfo{person}{Renzo Degiovanni}, {and} \bibinfo{person}{Mike Papadakis}.} \bibinfo{year}{2024}\natexlab{}.
\newblock \showarticletitle{Do {LLMs} generate test oracles that capture the actual or the expected program behaviour?}
\newblock \bibinfo{journal}{\emph{arXiv preprint arXiv:2410.21136}} (\bibinfo{year}{2024}).
\newblock


\bibitem[Kopetz and Steiner(2022)]%
        {kopetz:real}
\bibfield{author}{\bibinfo{person}{Hermann Kopetz} {and} \bibinfo{person}{Wilfried Steiner}.} \bibinfo{year}{2022}\natexlab{}.
\newblock \showarticletitle{Real-Time Communication}.
\newblock In \bibinfo{booktitle}{\emph{Real-time systems: Design principles for distributed embedded applications}}. \bibinfo{publisher}{Springer}, \bibinfo{pages}{177--200}.
\newblock


\bibitem[Lakhotia et~al\mbox{.}(2013)]%
        {kletal:austin-ist}
\bibfield{author}{\bibinfo{person}{Kiran Lakhotia}, \bibinfo{person}{Mark Harman}, {and} \bibinfo{person}{Hamilton Gross}.} \bibinfo{year}{2013}\natexlab{}.
\newblock \showarticletitle{{AUSTIN}: An Open Source Tool for Search Based Software Testing of {C} Programs}.
\newblock \bibinfo{journal}{\emph{Journal of Information and Software Technology}} \bibinfo{volume}{55}, \bibinfo{number}{1} (\bibinfo{date}{January} \bibinfo{year}{2013}), \bibinfo{pages}{112--125}.
\newblock


\bibitem[Luo et~al\mbox{.}(2014)]%
        {luo:flaky}
\bibfield{author}{\bibinfo{person}{Qingzhou Luo}, \bibinfo{person}{Farah Hariri}, \bibinfo{person}{Lamyaa Eloussi}, {and} \bibinfo{person}{Darko Marinov}.} \bibinfo{year}{2014}\natexlab{}.
\newblock \showarticletitle{An empirical analysis of flaky tests}. In \bibinfo{booktitle}{\emph{$22^{nd}$ International Symposium on Foundations of Software Engineering ({FSE 2014})}}, \bibfield{editor}{\bibinfo{person}{Shing-Chi Cheung}, \bibinfo{person}{Alessandro Orso}, {and} \bibinfo{person}{Margaret-Anne Storey}} (Eds.). \bibinfo{publisher}{ACM}, \bibinfo{address}{Hong Kong, China}, \bibinfo{pages}{643--653}.
\newblock
\showISBNx{978-1-4503-3056-5}


\bibitem[Madeyski et~al\mbox{.}(2013)]%
        {madeyski2013overcoming}
\bibfield{author}{\bibinfo{person}{Lech Madeyski}, \bibinfo{person}{Wojciech Orzeszyna}, \bibinfo{person}{Richard Torkar}, {and} \bibinfo{person}{Mariusz Jozala}.} \bibinfo{year}{2013}\natexlab{}.
\newblock \showarticletitle{Overcoming the equivalent mutant problem: A systematic literature review and a comparative experiment of second order mutation}.
\newblock \bibinfo{journal}{\emph{IEEE Transactions on Software Engineering}} \bibinfo{volume}{40}, \bibinfo{number}{1} (\bibinfo{year}{2013}), \bibinfo{pages}{23--42}.
\newblock


\bibitem[Man{\`{e}}s et~al\mbox{.}(2018)]%
        {manes:fuzzing}
\bibfield{author}{\bibinfo{person}{Valentin J.~M. Man{\`{e}}s}, \bibinfo{person}{HyungSeok Han}, \bibinfo{person}{Choongwoo Han}, \bibinfo{person}{Sang~Kil Cha}, \bibinfo{person}{Manuel Egele}, \bibinfo{person}{Edward~J. Schwartz}, {and} \bibinfo{person}{Maverick Woo}.} \bibinfo{year}{2018}\natexlab{}.
\newblock \showarticletitle{The Art, Science, and Engineering of Fuzzing: {A} Survey}.
\newblock \bibinfo{journal}{\emph{CoRR}}  \bibinfo{volume}{abs/1812.00140} (\bibinfo{year}{2018}).
\newblock
\showeprint[arxiv]{1812.00140}


\bibitem[Mao et~al\mbox{.}(2016)]%
        {mao:sapienz:16}
\bibfield{author}{\bibinfo{person}{Ke Mao}, \bibinfo{person}{Mark Harman}, {and} \bibinfo{person}{Yue Jia}.} \bibinfo{year}{2016}\natexlab{}.
\newblock \showarticletitle{Sapienz: Multi-objective Automated Testing for {Android} Applications}. In \bibinfo{booktitle}{\emph{International Symposium on Software Testing and Analysis ({ISSTA 2016})}}. \bibinfo{pages}{94--105}.
\newblock


\bibitem[Mao et~al\mbox{.}(2022)]%
        {kmetal:fausta}
\bibfield{author}{\bibinfo{person}{Ke Mao}, \bibinfo{person}{Timotej Kapus}, \bibinfo{person}{Lambros Petrou}, \bibinfo{person}{{\'{A}}kos Hajdu}, \bibinfo{person}{Matteo Marescotti}, \bibinfo{person}{Andreas L{\"{o}}scher}, \bibinfo{person}{Mark Harman}, {and} \bibinfo{person}{Dino Distefano}.} \bibinfo{year}{2022}\natexlab{}.
\newblock \showarticletitle{{FAUSTA:} Scaling Dynamic Analysis with Traffic Generation at WhatsApp}. In \bibinfo{booktitle}{\emph{15th {IEEE} Conference on Software Testing, Verification and Validation, {ICST} 2022, Valencia, Spain, April 4-14, 2022}}. \bibinfo{publisher}{{IEEE}}, \bibinfo{pages}{267--278}.
\newblock


\bibitem[McGill(2025)]%
        {mcgill:time}
\bibfield{author}{\bibinfo{person}{James McGill}.} \bibinfo{year}{2025}\natexlab{}.
\newblock \bibinfo{title}{Time to First Review}.
\newblock
\newblock
\urldef\tempurl%
\url{https://docs.velocity.codeclimate.com/en/articles/2913584-time-to-first-review}
\showURL{%
\tempurl}


\bibitem[{McMinn}(2004)]%
        {mcminn:survey}
\bibfield{author}{\bibinfo{person}{Phil {McMinn}}.} \bibinfo{year}{2004}\natexlab{}.
\newblock \showarticletitle{Search-based Software Test Data Generation: A Survey}.
\newblock \bibinfo{journal}{\emph{Software Testing, Verification and Reliability}} \bibinfo{volume}{14}, \bibinfo{number}{2} (\bibinfo{date}{June} \bibinfo{year}{2004}), \bibinfo{pages}{105--156}.
\newblock


\bibitem[Molina et~al\mbox{.}(2024)]%
        {molina:test}
\bibfield{author}{\bibinfo{person}{Facundo Molina}, \bibinfo{person}{Alessandra Gorla}, {and} \bibinfo{person}{Marcelo d’Amorim}.} \bibinfo{year}{2024}\natexlab{}.
\newblock \showarticletitle{Test Oracle Automation in the era of {LLMs}}.
\newblock \bibinfo{journal}{\emph{{ACM} Transactions on Software Engineering and Methodology}} (\bibinfo{year}{2024}).
\newblock


\bibitem[Mossige et~al\mbox{.}(2015)]%
        {mossige2015testing}
\bibfield{author}{\bibinfo{person}{Morten Mossige}, \bibinfo{person}{Arnaud Gotlieb}, {and} \bibinfo{person}{Hein Meling}.} \bibinfo{year}{2015}\natexlab{}.
\newblock \showarticletitle{Testing robot controllers using constraint programming and continuous integration}.
\newblock \bibinfo{journal}{\emph{Information and Software Technology}}  \bibinfo{volume}{57} (\bibinfo{year}{2015}), \bibinfo{pages}{169--185}.
\newblock


\bibitem[Murali et~al\mbox{.}(2024)]%
        {murali:codecompose}
\bibfield{author}{\bibinfo{person}{Vijayaraghavan Murali}, \bibinfo{person}{Chandra Maddila}, \bibinfo{person}{Imad Ahmad}, \bibinfo{person}{Michael Bolin}, \bibinfo{person}{Daniel Cheng}, \bibinfo{person}{Negar Ghorbani}, \bibinfo{person}{Renuka Fernandez}, \bibinfo{person}{Nachiappan Nagappan}, {and} \bibinfo{person}{Peter~C Rigby}.} \bibinfo{year}{2024}\natexlab{}.
\newblock \showarticletitle{CodeCompose: {A} Large-Scale Industrial Deployment of {AI}-assisted Code Authoring}. In \bibinfo{booktitle}{\emph{Foundations of Software Engineering ({FSE} 2024)}}. \bibinfo{address}{Porto de Galinhas, Brazil}.
\newblock
\newblock
\shownote{Earlier version available as arXiv:2305.12050}.


\bibitem[O'Hearn(2019)]%
        {poh:incorrectness}
\bibfield{author}{\bibinfo{person}{Peter~W O'Hearn}.} \bibinfo{year}{2019}\natexlab{}.
\newblock \showarticletitle{Incorrectness logic}.
\newblock \bibinfo{journal}{\emph{Proceedings of the ACM on Programming Languages}} \bibinfo{volume}{4}, \bibinfo{number}{POPL} (\bibinfo{year}{2019}), \bibinfo{pages}{1--32}.
\newblock


\bibitem[Papadakis et~al\mbox{.}(2015)]%
        {papadakis:trivial}
\bibfield{author}{\bibinfo{person}{Mike Papadakis}, \bibinfo{person}{Yue Jia}, \bibinfo{person}{Mark Harman}, {and} \bibinfo{person}{Yves~Le Traon}.} \bibinfo{year}{2015}\natexlab{}.
\newblock \showarticletitle{Trivial Compiler Equivalence: A Large Scale Empirical Study of a Simple, Fast and Effective Equivalent Mutant Detection Technique}. In \bibinfo{booktitle}{\emph{$37^{th}$ International Conference on Software Engineering ({ICSE 2015})}}. \bibinfo{address}{Florence, Italy}, \bibinfo{pages}{936--946}.
\newblock


\bibitem[Petke et~al\mbox{.}(2018)]%
        {Petke:gisurvey}
\bibfield{author}{\bibinfo{person}{Justyna Petke}, \bibinfo{person}{Saemundur~O. Haraldsson}, \bibinfo{person}{Mark Harman}, \bibinfo{person}{William~B. Langdon}, \bibinfo{person}{David~R. White}, {and} \bibinfo{person}{John~R. Woodward}.} \bibinfo{year}{2018}\natexlab{}.
\newblock \showarticletitle{Genetic Improvement of Software: a Comprehensive Survey}.
\newblock \bibinfo{journal}{\emph{IEEE Transactions on Evolutionary Computation}} \bibinfo{volume}{22}, \bibinfo{number}{3} (\bibinfo{date}{June} \bibinfo{year}{2018}), \bibinfo{pages}{415--432}.
\newblock


\bibitem[Pizzorno and Berger(2024)]%
        {pizzorno:coverup}
\bibfield{author}{\bibinfo{person}{Juan~Altmayer Pizzorno} {and} \bibinfo{person}{Emery~D Berger}.} \bibinfo{year}{2024}\natexlab{}.
\newblock \showarticletitle{Coverup: Coverage-guided {LLM}-based test generation}.
\newblock \bibinfo{journal}{\emph{arXiv preprint arXiv:2403.16218}} (\bibinfo{year}{2024}).
\newblock


\bibitem[R\"{a}ih\"{a}(2010)]%
        {raiha:survey}
\bibfield{author}{\bibinfo{person}{Outi R\"{a}ih\"{a}}.} \bibinfo{year}{2010}\natexlab{}.
\newblock \showarticletitle{A survey on Search--Based Software Design}.
\newblock \bibinfo{journal}{\emph{Computer Science Review}} \bibinfo{volume}{4}, \bibinfo{number}{4} (\bibinfo{year}{2010}), \bibinfo{pages}{203--249}.
\newblock


\bibitem[Ramirez et~al\mbox{.}(2018)]%
        {ramirez2018systematic}
\bibfield{author}{\bibinfo{person}{Aurora Ramirez}, \bibinfo{person}{Jose~Raul Romero}, {and} \bibinfo{person}{Christopher~L Simons}.} \bibinfo{year}{2018}\natexlab{}.
\newblock \showarticletitle{A systematic review of interaction in search-based software engineering}.
\newblock \bibinfo{journal}{\emph{IEEE Transactions on Software Engineering}} \bibinfo{volume}{45}, \bibinfo{number}{8} (\bibinfo{year}{2018}), \bibinfo{pages}{760--781}.
\newblock


\bibitem[Ramirez et~al\mbox{.}(2019)]%
        {ramirez2019survey}
\bibfield{author}{\bibinfo{person}{Aurora Ramirez}, \bibinfo{person}{Jos{\'e}~Ra{\'u}l Romero}, {and} \bibinfo{person}{Sebastian Ventura}.} \bibinfo{year}{2019}\natexlab{}.
\newblock \showarticletitle{A survey of many-objective optimisation in search-based software engineering}.
\newblock \bibinfo{journal}{\emph{Journal of Systems and Software}}  \bibinfo{volume}{149} (\bibinfo{year}{2019}), \bibinfo{pages}{382--395}.
\newblock


\bibitem[Schuler and Zeller(2009)]%
        {schuler:javalanche}
\bibfield{author}{\bibinfo{person}{David Schuler} {and} \bibinfo{person}{Andreas Zeller}.} \bibinfo{year}{2009}\natexlab{}.
\newblock \showarticletitle{Javalanche: efficient mutation testing for {J}ava}. In \bibinfo{booktitle}{\emph{$7^{th}$ joint meeting of the European Software Engineering Conference and the {ACM SIGSOFT} International Symposium on Foundations of Software Engineering ({ESEC/FSE 2009})}}. \bibinfo{pages}{297--298}.
\newblock


\bibitem[Sen et~al\mbox{.}(2005)]%
        {sen:cute}
\bibfield{author}{\bibinfo{person}{Koushik Sen}, \bibinfo{person}{Darko Marinov}, {and} \bibinfo{person}{Gul Agha}.} \bibinfo{year}{2005}\natexlab{}.
\newblock \showarticletitle{{CUTE}: a concolic unit testing engine for {C}}. In \bibinfo{booktitle}{\emph{$10^{th}$ European Software Engineering Conference and 13th {ACM} International Symposium on Foundations of Software Engineering ({ESEC/FSE} '05)}}, \bibfield{editor}{\bibinfo{person}{Michel Wermelinger} {and} \bibinfo{person}{Harald Gall}} (Eds.). \bibinfo{publisher}{ACM}, \bibinfo{pages}{263--272}.
\newblock
\showISBNx{1-59593-014-0}


\bibitem[Simons et~al\mbox{.}(2010)]%
        {simons:interactive}
\bibfield{author}{\bibinfo{person}{Christopher~L. Simons}, \bibinfo{person}{Ian~C. Parmee}, {and} \bibinfo{person}{Rhys Gwynllyw}.} \bibinfo{year}{2010}\natexlab{}.
\newblock \showarticletitle{Interactive, Evolutionary Search in Upstream Object-Oriented Class Design}.
\newblock \bibinfo{journal}{\emph{{IEEE} Transactions on Software Engineering}} \bibinfo{volume}{36}, \bibinfo{number}{6} (\bibinfo{year}{2010}), \bibinfo{pages}{798--816}.
\newblock


\bibitem[Steven et~al\mbox{.}(2000)]%
        {steven:jrapture}
\bibfield{author}{\bibinfo{person}{John Steven}, \bibinfo{person}{Pravir Chandra}, \bibinfo{person}{Bob Fleck}, {and} \bibinfo{person}{Andy Podgurski}.} \bibinfo{year}{2000}\natexlab{}.
\newblock \showarticletitle{{jRapture}: {A} capture/replay tool for observation-based testing}. In \bibinfo{booktitle}{\emph{Proceedings of the 2000 {ACM SIGSOFT} international symposium on Software Testing and Analysis ({ISSTA 2000})}}. \bibinfo{pages}{158--167}.
\newblock


\bibitem[Tiwari et~al\mbox{.}(2023)]%
        {tiwari:mimicking}
\bibfield{author}{\bibinfo{person}{Deepika Tiwari}, \bibinfo{person}{Martin Monperrus}, {and} \bibinfo{person}{Benoit Baudry}.} \bibinfo{year}{2023}\natexlab{}.
\newblock \bibinfo{title}{Mimicking Production Behavior with Generated Mocks}.
\newblock
\newblock
\showeprint[arxiv]{2208.01321}~[cs.SE]


\bibitem[Tuli et~al\mbox{.}(2023)]%
        {tuli:simulation}
\bibfield{author}{\bibinfo{person}{Shreshth Tuli}, \bibinfo{person}{Kinga Bojarczuk}, \bibinfo{person}{Natalija Gucevska}, \bibinfo{person}{Mark Harman}, \bibinfo{person}{Xiao{-}Yu Wang}, {and} \bibinfo{person}{Graham Wright}.} \bibinfo{year}{2023}\natexlab{}.
\newblock \showarticletitle{Simulation-Driven Automated End-to-End Test and Oracle Inference}. In \bibinfo{booktitle}{\emph{45th {IEEE/ACM} International Conference on Software Engineering: Software Engineering in Practice, SEIP@ICSE 2023, Melbourne, Australia, May 14-20, 2023}}. \bibinfo{publisher}{{IEEE}}, \bibinfo{pages}{122--133}.
\newblock


\bibitem[van Hijfte and Oprescu(2021)]%
        {van2021mutantbench}
\bibfield{author}{\bibinfo{person}{Lars van Hijfte} {and} \bibinfo{person}{Ana Oprescu}.} \bibinfo{year}{2021}\natexlab{}.
\newblock \showarticletitle{Mutantbench: an equivalent mutant problem comparison framework}. In \bibinfo{booktitle}{\emph{2021 IEEE International Conference on Software Testing, Verification and Validation Workshops (ICSTW)}}. IEEE, \bibinfo{pages}{7--12}.
\newblock


\bibitem[Veanes et~al\mbox{.}(2005)]%
        {veanes2005fly}
\bibfield{author}{\bibinfo{person}{Margus Veanes}, \bibinfo{person}{Colin Campbell}, \bibinfo{person}{Wolfram Schulte}, \bibinfo{person}{Pushmeet Kohli}, \bibinfo{person}{N Tillmann}, {and} \bibinfo{person}{W Grieskamp}.} \bibinfo{year}{2005}\natexlab{}.
\newblock \showarticletitle{On-the-fly testing of reactive systems}.
\newblock \bibinfo{journal}{\emph{Submitted for publication}} (\bibinfo{year}{2005}).
\newblock


\bibitem[Wang et~al\mbox{.}(2023)]%
        {wang:LLM-testing-survey}
\bibfield{author}{\bibinfo{person}{Junjie Wang}, \bibinfo{person}{Yuchao Huang}, \bibinfo{person}{Chunyang Chen}, \bibinfo{person}{Zhe Liu}, \bibinfo{person}{Song Wang}, {and} \bibinfo{person}{Qing Wang}.} \bibinfo{year}{2023}\natexlab{}.
\newblock \showarticletitle{Software testing with large language model: Survey, landscape, and vision}.
\newblock \bibinfo{journal}{\emph{arXiv preprint arXiv:2307.07221}} (\bibinfo{year}{2023}).
\newblock


\bibitem[Wang et~al\mbox{.}(2024)]%
        {wang:hits}
\bibfield{author}{\bibinfo{person}{Zejun Wang}, \bibinfo{person}{Kaibo Liu}, \bibinfo{person}{Ge Li}, {and} \bibinfo{person}{Zhi Jin}.} \bibinfo{year}{2024}\natexlab{}.
\newblock \showarticletitle{{HITS}: High-coverage {LLM}-based Unit Test Generation via Method Slicing}. In \bibinfo{booktitle}{\emph{Proceedings of the 39th IEEE/ACM International Conference on Automated Software Engineering}}. \bibinfo{pages}{1258--1268}.
\newblock


\bibitem[Zalewski(2025)]%
        {zalewski:afl}
\bibfield{author}{\bibinfo{person}{Michal Zalewski}.} \bibinfo{year}{Accessed March 27th 2025}\natexlab{}.
\newblock \bibinfo{title}{American fuzzy lop}.
\newblock
\newblock
\urldef\tempurl%
\url{http://lcamtuf.coredump.cx/afl/}
\showURL{%
\tempurl}


\bibitem[Zhao(2023)]%
        {copilot2023}
\bibfield{author}{\bibinfo{person}{Shuyin Zhao}.} \bibinfo{year}{2023}\natexlab{}.
\newblock \bibinfo{title}{GitHub {C}opilot now has a better {AI} model and new capabilities}.
\newblock
\newblock
\urldef\tempurl%
\url{https://github.blog/2023-02-14-github-copilot-now-has-a-better-ai-model-and-new-capabilities/}
\showURL{%
\tempurl}


\bibitem[Zhu and Pan(2019)]%
        {zhu2019automatic}
\bibfield{author}{\bibinfo{person}{Yuxiang Zhu} {and} \bibinfo{person}{Minxue Pan}.} \bibinfo{year}{2019}\natexlab{}.
\newblock \showarticletitle{Automatic code summarization: A systematic literature review}.
\newblock \bibinfo{journal}{\emph{arXiv preprint arXiv:1909.04352}} (\bibinfo{year}{2019}).
\newblock


\end{thebibliography}

\end{document}